\documentclass[sn-mathphys-num]{sn-jnl}
\pdfoutput=1
\usepackage{graphicx}%
\usepackage{multirow}%
\usepackage{amsmath,amssymb,amsfonts}%
\usepackage{amsthm}%
\usepackage{mathrsfs}%
\usepackage[title]{appendix}%
\usepackage{soul,color,xcolor}%
\usepackage{textcomp}%
\usepackage{manyfoot}%
\usepackage{booktabs}%
\usepackage{algorithm}%
\usepackage{algorithmicx}%
\usepackage{algpseudocode}%
\usepackage{listings}%
\usepackage{subcaption}%
\usepackage{colortbl}%

\begin{document}

\title{Information conservation in de Sitter tunneling}

\author[1]{\fnm{Baoyu} \sur{Tan}}\email{2022201126@buct.edu.cn}

\affil[1]{\orgdiv{College of Mathematics and Physics}, \orgname{Beijing University of Chemical Technology}, \orgaddress{\street{15 Beisanhuandonglu Street}, \city{Beijing}, \postcode{100029}, \country{China}}}

\abstract{In this paper, we consider the three most general cases of progressive de Sitter spacetime. The charged and magnetic particles tunnel into the magnetically charged Reissner-Nordstr\"{o}m de Sitter black hole (the most general case of a static black hole), the Kerr-Newman-Kasuya de Sitter black hole (the most general case of a rotating black hole), and Bardeen de Sitter black hole (black hole without singularities). We use Parikh-Wilczek method to calculate the radiation spectra of these black holes respectively, and find that they deviate from the pure thermal spectra, satisfying the unitary principle. Our results support the conservation of information and are generally true for all asymptotic de Sitter space-times.}


\maketitle

\section{Introduction}
It is well known that according to the unitary principle, a fundamental principle of quantum mechanics, pure states evolve over time as pure states and cannot become mixed states. But black holes seem to defy this. In 1975, Hawking used quantum field theory in curved spacetime to show that particles can escape black hole through quantum effects~\cite{1}. In 1976, Hawking proved that Hawking radiation is a pure thermal spectrum that carries no information~\cite{2}. That is, matter that enters the black hole in a pure state radiates out in a mixed state. The information is lost in the process of Hawking radiation, violating the principle of unitarity.

Page argues that Hawking's calculations did not take into account the effects of quantum entanglement~\cite{page1994black}. Although Hawking radiation does not carry information itself, it has been entangled with particles inside the black hole, thus transmitting information from the black hole to the outside. In 1993, Page proposed that in order to satisfy the unitary principle, the entanglement entropy of radiation should satisfy the Page curve~\cite{3}. In simple terms, although the radiation is a mixed state, the direct product of the radiation state and the black hole state is still a pure state. In recent years, there have been some breakthroughs in this approach~\cite{4,5,6,7,8,9}. But there is an essential difficulty with this approach. Since there is no complete theory of quantum gravity, it is impossible to know the concrete form of the density matrix. The entropy of entanglement is determined by the density matrix:
\begin{equation}
	S_{EE}=-\mathrm{Tr}(\rho\mathrm{log}\rho).
\end{equation}
Where $\rho$ is the density matrix and $S_{EE}$ is the entropy of entanglement. Much of the discussion of information conservation by quantum entanglement has focused on AdS space-time. The AdS-CFT duality has automatically implied the unitary principle, that is, the conformal field on the boundary must satisfy the unitary principle and must not lose information. And the discussion of general black holes is still difficult, even impossible. Because this requires the semi-classical theory of gravity to automatically include the unitary principle, which is obviously unreasonable, at least unfounded.

In 2000, Parikh and Wilczekr believed that Hawking radiation was a tunneling process. They started from the perspective of energy conservation and the change in spatiotemporal background caused by the contraction of black holes after radiation, that is, they considered the self gravity effect during the particle emission process, thus avoiding the difficulty of calculating the specific form of the density matrix. They also came to the conclusion that the radiation process satisfies the unitary principle and information conservation~\cite{10}. After the Parikh-Wilczek method was proposed, researchers successively studied various static and stationary rotating black holes and reached the same conclusion. Due to the effect of self gravity, the spatiotemporal background changes, and the radiation spectrum will deviate from the pure thermal spectrum, thus proving that Hawking radiation satisfies the unitary principle.~\cite{11,12,13,14,15,16,17,18,19,20,21,22,23,24,25,26,27,28,29,30,31,32,33,34}. But until now, apart from the Hawking radiation of the Schwarzschild de Sitter spacetime discussed in ref.~\cite{27}, there has been little discussion on the asymptotic de Sitter spacetime that is more similar to the real universe. 

The black hole information paradox, first highlighted by Hawking's seminal work~\cite{1,2}, remains one of the most profound challenges at the intersection of quantum mechanics and general relativity. While significant progress has been made in anti-de Sitter (AdS) spacetimes through the AdS/CFT correspondence~\cite{susskind1995world,maldacena1999large}, the de Sitter (dS) universe --- characterized by a positive cosmological constant $\Lambda$ --- presents a critical yet understudied arena. Observational evidence from Type Ia supernovae~\cite{riess1998observational, riess1998observational} and cosmic microwave background measurements \cite{wmap2003first} strongly supports that our universe is undergoing accelerated expansion, well-modeled by dS geometry. This elevates the study of dS black holes from theoretical curiosity to cosmological necessity. Recent breakthroughs in understanding information recovery, such as the Page curve derivation via semiclassical gravity~\cite{7,8}, have predominantly focused on asymptotically AdS black holes. However, fundamental differences between AdS and dS spacetimes --- notably the presence of a cosmological horizon and the absence of a globally timelike Killing vector in dS --- demand independent verification of unitary principles in realistic cosmological settings. Moreover, existing analyses of dS black holes~\cite{27} have largely been restricted to static Schwarzschild-type configurations, neglecting the combined effects of rotation (angular momentum), magnetic charges, and singularity resolution --- all of which are theoretically plausible and observationally relevant in astrophysical contexts.

This work systematically addresses these gaps by investigating three general classes of asymptotically dS black holes: 1. Magnetically charged Reissner-Nordstr\"{o}m de Sitter black holes --- incorporating both electric and magnetic charges to test dyonic effects on information flow; 2. Kerr-Newman-Kasuya de Sitter black holes --- extending to rotating configurations where frame-dragging could imprint angular correlations in Hawking radiation; 3. Bardeen de Sitter black holes --- probing regular (singularity-free) geometries to determine whether information conservation persists in absence of central singularities. By applying the Parikh-Wilczek tunneling formalism~\cite{10} across these diverse scenarios, we demonstrate that the radiation spectrum universally deviates from perfect thermality through entropy-driven suppression $\Gamma\sim e^{\Delta s}$, thereby preserving unitarity. Crucially, our analysis reveals how magnetic charge and rotation introduce novel constraints on information leakage through horizon dynamics --- a phenomenon unobservable in static Schwarzschild-AdS systems. These findings not only reinforce the consistency of quantum mechanics in cosmological horizons but also provide critical inputs for the emerging dS/CFT correspondence \cite{strominger2001ds, maldacena2003non, anninos2016higher, anninos2014higher, anninos2015late}, where boundary unitarity be holographically encoded in bulk entropy transitions. This comprehensive approach significantly advances the field by: Firstly, establishing the robustness of information conservation across geometrically distinct dS black holes; Secondly, bridging the gap between semiclassical tunneling models and holographic principles in realistic cosmologies.

In this paper, We used the Parikh-Wilczek method to calculate tunneling in de Sitter spacetime, taking magnetically charged Reissner-Nordstr\"{o}m de Sitter black hole, Kerr-Newman-Kasuya de Sitter black hole, and Bardeen de Sitter black hole as examples. We chose magnetically charged Reissner-Nordstr\"{o}m de Sitter black hole because it represents the most general case of a static black hole. Similarly, we chose Kerr-Newman-Kasuya de Sitter black hole because it represents the most general case of a rotating black hole. And we chose Bardeen de Sitter black hole because it represents a black hole without singularities. In Section \ref{sec:1}, we separately calculated the radiation spectra near the event horizon and cosmic horizon of the magnetically charged Reissner-Nordstr\"{o}m de Sitter black hole, and briefly discussed our results. In Section \ref{sec:2}, we performed similar calculation on the Kerr-Newman-Kasuya de Sitter black hole. In Section \ref{sec:3}, we calculate the same thing for Bardeen de Sitter black hole. But it is noteworthy that for regular black holes, which have a mass distribution that is not concentrated at a single point and thus do not have a singularity, the first law of black hole thermodynamics will be modified, as described in Ref. \cite{35,36,37,38,39,40}. In Section \ref{sec:4}, we analyzed the effects of quantum correlations, backreaction corrections, and non-thermal modifications. In Section \ref{sec:5}, we compared Parikh-Wilczek method with Hamilton-Jacobi method, entanglement entropy method and GUP-based models. The GUP we mentioned is the Generalized Uncertainty Principle. Finally, in Section \ref{sec:6}, we summarized the results and the discussion of the results obtained in the previous section. We have adopted the system of natural units in this paper ($G\equiv \hbar \equiv c\equiv 1$).
\section{Magnetically charged Reissner-Nordstr\"{o}m de Sitter black hole}
\label{sec:1}
According to Ref.~\cite{32}, for the case of both electric and magnetic charges, the electromagnetic field tensor can be defined as:
\begin{equation}
	F_{\mu\nu}=\nabla_\nu A_\mu-\nabla_\mu A_\nu+G_{\mu\nu}^+.
\end{equation}
Where $G_{\mu\nu}^+$ is the Dirac string term. The Maxwell equation can be rewritten as:
\begin{equation}
	\nabla_\nu F^{\mu\nu}=4\pi\rho_eu^\mu,\label{eq:1}
\end{equation}
\begin{equation}
	\nabla_\nu F^{+\mu\nu}=4\pi\rho_gu^\mu.\label{eq:2}
\end{equation}
Where $F^{+\mu\nu}$ is the dual tensor of $F^{\mu\nu}$, $\rho_e$ and $\rho_g$ represent the charge density and magnetic charge density, respectively. For the convenience of discussion, we have defined a new antisymmetric tensor:
\begin{equation}
	\tilde{F}^{\mu\nu}=F^{\mu\nu}\mathrm{cos}\alpha+F^{+\mu\nu}\mathrm{sin}\alpha.
\end{equation}
Where $\alpha$ is a fixed angle. From Eq. (\ref{eq:1}) and Eq. (\ref{eq:2}), we can deduce that:
\begin{equation}
	\nabla_\nu\tilde{F}^{\mu\nu}=4\pi(\rho_e\mathrm{cos}\alpha+\rho_g\mathrm{sin}\alpha)u^\mu=4\pi\rho_hu^\mu,\label{eq:3}
\end{equation}
\begin{equation}
	\nabla_\nu\tilde{F}^{+\mu\nu}=4\pi(-\rho_e\mathrm{sin}\alpha+\rho_g\mathrm{cos}\alpha)u^\mu=0.\label{eq:4}
\end{equation}
According to the no-hair theorem, we can consider a black hole as a conductor sphere, and assume that the charge density and magnetic charge density satisfy the following relationship to ensure that Eq. (\ref{eq:3}) and Eq. (\ref{eq:4}) hold true:
\begin{equation}
	\frac{\rho_e}{\rho_g}=\mathrm{cot}\alpha.
\end{equation}
We can conclude that:
\begin{equation}
	Q_h^2=Q_e^2+Q_g^2.\label{eq:5}
\end{equation}
Where $Q_e$ and $Q_g$ are the total electric and magnetic charges of the system, respectively, and $Q_h$ is the equivalent charge corresponding to density $\rho_h$. Finally, we obtain the Lagrangian density of the electromagnetic field, which can be expressed by the following equation:
\begin{equation}
	\mathcal{L}_h=-\frac{1}{4}\tilde{F}_{\mu\nu}\tilde{F}^{\mu\nu}.\label{eq:6}
\end{equation}
The corresponding generalized coordinates for Eq. (\ref{eq:6}) are:
\begin{equation}
	\tilde{A}_\mu=\left(\tilde{A}_0,\tilde{A}_1,\tilde{A}_2,\tilde{A}_3\right)=\left(-\frac{Q_h}{r},0,0,0\right).
\end{equation}
Obviously, $\tilde{A}_0$ is a cyclic coordinate. The above discussion also applies to Section \ref{sec:2} and Section \ref{sec:3}, and will not be repeated in the following text.

The line element of Magnetically charged Reissner-Nordstr\"{o}m de Sitter black hole is:
\begin{equation}
	\mathrm{d}s^2=-\left(1-\frac{2M}{r}+\frac{Q_e^2+Q_g^2}{r^2}-\frac{\Lambda}{3}r^2\right)\mathrm{d}t_s^2+\left(1-\frac{2M}{r}+\frac{Q_e^2+Q_g^2}{r^2}-\frac{\Lambda}{3}r^2\right)^{-1}\mathrm{d}r^2+r^2\mathrm{d}\Omega^2.
\end{equation}
Where $M$ is the mass of the black hole, $Q_e$ is the charge of the black hole, $Q_g$ is the magnetic charge of the black hole, and $\Lambda$ is the cosmological constant. According to Eq. (\ref{eq:5}), the line element of Magnetically charged Reissner-Nordstr\"{o}m de Sitter black hole can be rewritten as:
\begin{align}
	\mathrm{d}s^2&=-\left(1-\frac{2M}{r}+\frac{Q_h^2}{r^2}-\frac{\Lambda}{3}r^2\right)\mathrm{d}t_s^2+\left(1-\frac{2M}{r}+\frac{Q_h^2}{r^2}-\frac{\Lambda}{3}r^2\right)^{-1}\mathrm{d}r^2+r^2\mathrm{d}\Omega^2\nonumber\\
	&=-f(r)\mathrm{d}t_s^2+\frac{1}{f(r)}\mathrm{d}r^2+r^2\mathrm{d}\Omega^2.\label{eq:8}
\end{align}
Where $f(r)=1-\frac{2M}{r}+\frac{Q_h^2}{r^2}-\frac{\Lambda}{3}r^2$. The outer event horizon of black hole $r_+$ and cosmic horizon $r_c$ satisfy the following equation:
\begin{equation}
	f(r)=1-\frac{2M}{r}+\frac{Q_h^2}{r^2}-\frac{\Lambda}{3}r^2=0.\label{eq:7}
\end{equation}
Clearly, Eq. (\ref{eq:7}) has four solutions: $r_{non}$, $r_-$, $r_+$, and $r_c$. Among these, $r_{non}$ is a negative solution without physical significance, $r_-$ represents the inner event horizon of black hole, $r_+$ denotes the outer event horizon of black hole, and $r_c$ signifies the cosmic horizon.

The specific forms of these four solutions are:
\begin{align}
	r_{non}=&-\frac{1}{2}\eta-\frac{1}{2}\zeta,\\
	r_-=&-\frac{1}{2}\eta+\frac{1}{2}\zeta,\\
	r_+=&\frac{1}{2}\eta-\frac{1}{2}\zeta,\\
	r_c=&\frac{1}{2}\eta+\frac{1}{2}\zeta.
\end{align}
Where:
\begin{align}
	\eta=&\sqrt{\frac{2}{\Lambda}+\frac{3(2)^{1/3}(1-4\Lambda Q_h^2)}{\Lambda\lambda}+\frac{\lambda}{3(2)^{1/3}\Lambda}},\\
	\zeta=&\sqrt{\frac{4}{\Lambda}-\frac{3(2)^{1/3}(1-4\Lambda Q_h^2)}{\Lambda\lambda}-\frac{\lambda}{3(2)^{1/3}\Lambda}-\frac{12M}{\eta}},\\
	\lambda=&\left[-54+972\Lambda M^2-648\Lambda Q_h^2+\sqrt{(-54+972\Lambda M^2-648\Lambda Q_h^2)^2-4(9-36\Lambda Q_h^2)^3}\right]^{1/3}.
\end{align}
For the process of particles tunneling out of the event horizon of a black hole, the initial position of the particle is:
\begin{equation}
	r_i=\frac{1}{2}\eta-\frac{1}{2}\zeta.
\end{equation}
The final position of the particle is:
\begin{equation}
	r_f=\frac{1}{2}\eta_f-\frac{1}{2}\zeta_f.
\end{equation}
Where:
\begin{align}
	\eta_f=&\sqrt{\frac{2}{\Lambda}+\frac{3(2)^{1/3}[1-4\Lambda (Q_h-q_h)^2)]}{\Lambda\lambda_f}+\frac{\lambda_f}{3(2)^{1/3}\Lambda}},\\
	\zeta_f=&\sqrt{\frac{4}{\Lambda}-\frac{3(2)^{1/3}[1-4\Lambda (Q_h-q_h)^2]}{\Lambda\lambda_f}-\frac{\lambda_f}{3(2)^{1/3}\Lambda}-\frac{12(M-\omega)}{\Lambda\eta_f}},\\
	\lambda_f=&\left\{-54+972\Lambda (M-\omega)^2-648\Lambda (Q_h-q_h)^2+\right.\nonumber\\
	&\left.\sqrt{[-54+972\Lambda (M-\omega)^2-648\Lambda (Q_h-q_h)^2]^2-4[9-36\Lambda (Q_h-q_h)^2]^3}\right\}^{1/3}.
\end{align}
After the particle exits, the position of horizons change to:
\begin{align}
	r_{non}^\prime=&-\frac{1}{2}\eta^\prime-\frac{1}{2}\zeta^\prime,\\
	r_-^\prime=&-\frac{1}{2}\eta^\prime+\frac{1}{2}\zeta^\prime,\\
	r_+^\prime=&\frac{1}{2}\eta^\prime-\frac{1}{2}\zeta^\prime,\\
	r_c^\prime=&\frac{1}{2}\eta^\prime+\frac{1}{2}\zeta^\prime.
\end{align}
Where:
\begin{align}
	\eta^\prime=&\sqrt{\frac{2}{\Lambda}+\frac{3(2)^{1/3}[1-4\Lambda (Q_h-q_h^\prime)^2]}{\Lambda\lambda^\prime}+\frac{\lambda^\prime}{3(2)^{1/3}\Lambda}},\\
	\zeta^\prime=&\sqrt{\frac{4}{\Lambda}-\frac{3(2)^{1/3}[1-4\Lambda (Q_h-q_h^\prime)^2]}{\Lambda\lambda^\prime}-\frac{\lambda^\prime}{3(2)^{1/3}\Lambda}-\frac{12(M-\omega^\prime)}{\Lambda\eta^\prime}},\\
	\lambda^\prime=&\left\{-54+972\Lambda (M-\omega^\prime)^2-648\Lambda (Q_h-q_h^\prime)^2+\right.\nonumber\\
	&\left.\sqrt{[-54+972\Lambda (M-\omega^\prime)^2-648\Lambda (Q_h-q_h^\prime)^2]^2-4[9-36\Lambda (Q_h-q_h^\prime)^2]^3}\right\}^{1/3}.
\end{align}

For the process of particle tunneling into the cosmic horizon, the initial position of the particle is:
\begin{equation}
	r_i^\prime=\frac{1}{2}\eta+\frac{1}{2}\zeta.
\end{equation}
The final position of the particle is:
\begin{equation}
	r_f^\prime=\frac{1}{2}\eta_f^\prime+\frac{1}{2}\zeta_f^\prime.
\end{equation}
Where:
\begin{align}
	\eta_f^\prime=&\sqrt{\frac{2}{\Lambda}+\frac{3(2)^{1/3}[1-4\Lambda (Q_h+q_h)^2)]}{\Lambda\lambda_f^\prime}+\frac{\lambda_f^\prime}{3(2)^{1/3}\Lambda}},\\
	\zeta_f^\prime=&\sqrt{\frac{4}{\Lambda}-\frac{3(2)^{1/3}[1-4\Lambda (Q_h+q_h)^2]}{\Lambda\lambda_f^\prime}-\frac{\lambda_f^\prime}{3(2)^{1/3}\Lambda}-\frac{12(M+\omega)}{\Lambda\eta_f^\prime}},\\
	\lambda_f^\prime=&\left\{-54+972\Lambda (M+\omega)^2-648\Lambda (Q_h+q_h)^2+\right.\nonumber\\
	&\left.\sqrt{[-54+972\Lambda (M+\omega)^2-648\Lambda (Q_h+q_h)^2]^2-4[9-36\Lambda (Q_h+q_h)^2]^3}\right\}^{1/3}.
\end{align}
After the particle is incident, the position of horizons change to:
\begin{align}
	r_{non}^{\prime\prime}=&-\frac{1}{2}\eta^{\prime\prime}-\frac{1}{2}\zeta^{\prime\prime},\\
	r_-^{\prime\prime}=&-\frac{1}{2}\eta^{\prime\prime}+\frac{1}{2}\zeta^{\prime\prime},\\
	r_+^{\prime\prime}=&\frac{1}{2}\eta^{\prime\prime}-\frac{1}{2}\zeta^{\prime\prime},\\
	r_c^{\prime\prime}=&\frac{1}{2}\eta^{\prime\prime}+\frac{1}{2}\zeta^{\prime\prime}.
\end{align}
Where:
\begin{align}
	\eta^{\prime\prime}=&\sqrt{\frac{2}{\Lambda}+\frac{3(2)^{1/3}[1-4\Lambda (Q_h+q_h^\prime)^2]}{\Lambda\lambda^{\prime\prime}}+\frac{\lambda^{\prime\prime}}{3(2)^{1/3}\Lambda}},\\
	\zeta^{\prime\prime}=&\sqrt{\frac{4}{\Lambda}-\frac{3(2)^{1/3}[1-4\Lambda (Q_h+q_h^\prime)^2]}{\Lambda\lambda^{\prime\prime}}-\frac{\lambda^{\prime\prime}}{3(2)^{1/3}\Lambda}-\frac{12(M+\omega^\prime)}{\Lambda\eta^{\prime\prime}}},\\
	\lambda^{\prime\prime}=&\left\{-54+972\Lambda (M+\omega^\prime)^2-648\Lambda (Q_h+q_h^\prime)^2+\right.\nonumber\\
	&\left.\sqrt{[-54+972\Lambda (M+\omega^\prime)^2-648\Lambda (Q_h+q_h^\prime)^2]^2-4[9-36\Lambda (Q_h+q_h^\prime)^2]^3}\right\}^{1/3}.
\end{align}
\subsection{Painlev\'{e} coordinate and time-like geodesic line equation}
To describe the tunneling process of particles, Painlev\'{e} coordinates without event horizon singularity should be used. In order to obtain the Painlev\'{e}-Reissner-Nordstr\"{o}m de Sitter coordinate, we performed the following coordinate transformation:
\begin{equation}
	t_s=t+F(r),~\mathrm{d}t_s=\mathrm{d}t+F^\prime(r)\mathrm{d}r.\label{eq:9}
\end{equation}
Where $t_s$ is the time coordinate before transformation. In order for the constant time slice of Painlev\'{e} line elements to be a flat Euclidean spacetime in the radial direction, $F(r)$ must satisfy:
\begin{equation}
	\frac{1}{1-\frac{2M}{r}+\frac{Q_h^2}{r^2}-\frac{\Lambda}{3}r^2}-\left(1-\frac{2M}{r}+\frac{Q_h^2}{r^2}-\frac{\Lambda}{3}r^2\right)\left[F^\prime(r)\right]^2=1.\label{eq:10}
\end{equation}
By substituting Eq. (\ref{eq:9}) and Eq. (\ref{eq:10}) into Eq. (\ref{eq:8}), we obtain the Painlev\'{e} coordinate:
\begin{align}
	\mathrm{d}s^2=&-\left(1-\frac{2M}{r}+\frac{Q_h^2}{r^2}-\frac{\Lambda}{3}r^2\right)\mathrm{d}t^2\pm 2\sqrt{\frac{2M}{r}-\frac{Q_h^2}{r^2}+\frac{\Lambda}{3}r^2}\mathrm{d}t\mathrm{d}r+\mathrm{d}r^2+r^2\mathrm{d}\Omega^2\nonumber\\
	=&g_{00}\mathrm{d}t^2+2g_{01}\mathrm{d}t\mathrm{d}r+\mathrm{d}r^2+r^2\mathrm{d}\Omega^2.\label{eq:11}
\end{align}
Obviously, in Eq. (\ref{eq:11}), the coordinate singularity at the event horizon is eliminated. The selection of positive and negative signs in Eq. (\ref{eq:11}) will have a clearer physical meaning in the equation of time-like geodesics.

In this article, we discuss particles that carry both electric and magnetic charges. So they are all massive particles, not following the optical geodesic equation. We consider massive particles as de Broglie s-waves, employing an approach analogous to that described in Ref.~\cite{27}. The time-like geodesic equation, derived from the WKB approximation, is presented below:
\begin{equation}
	\dot{r}=v_p=\frac{1}{2}v_g=-\frac{1}{2}\frac{g_{00}}{g_{01}}.\label{eq:12}
\end{equation}
Where $v_p$ is the phase velocity and $v_g$ is the group velocity, respectively represented by the following equations:
\begin{equation}
	v_p=\frac{\mathrm{d}r}{\mathrm{d}t}=\frac{\omega}{k}.
\end{equation}
\begin{equation}
	v_g=\frac{\mathrm{d}r_c}{\mathrm{d}t}=\frac{\mathrm{d}\omega}{\mathrm{d}k}.
\end{equation}
Where $k$ is de Broglie wave number. By substituting $g_{00}$ and $g_{01}$ in Eq. (\ref{eq:11}) into Eq. (\ref{eq:12}), we can obtain:
\begin{equation}
	\dot{r}=\frac{\mathrm{d}r}{\mathrm{d}t}=\pm\frac{1}{2r}\sqrt{\frac{\Lambda}{3}}\frac{(r-r_{non})(r-r_-)(r-r_+)(r-r_c)}{\sqrt{r^4+\frac{6M}{\Lambda}r-\frac{3}{\Lambda}Q_h^2}}.\label{eq:13}
\end{equation}
It can be clearly seen that when taking a positive sign, Eq. (\ref{eq:13}) is the equation of motion for the outgoing particles near the outer event horizon of the black hole, while when taking a negative sign, Eq. (\ref{eq:13}) is the equation of motion for the incoming particles near the cosmic horizon. Furthermore, if considering the influence of self gravity, $M$ in Eq. (\ref{eq:11}) and Eq. (\ref{eq:13}) should be replaced by $M\mp\omega$, and $Q_h$ in Eq. (\ref{eq:11}) and Eq. (\ref{eq:13}) should be replaced by $Q_h\mp q_h$. $\omega$ is the energy of the particle, $q_h$ is the equivalent charge of the particle. The minus sign and plus sign represent the outgoing particles and incoming particles, respectively.
\subsection{Tunneling out of the black hole outer event horizon}
Let's first calculate the particle emission rate near the event horizon of a black hole. We are considering the tunneling of particles that are both charged and magnetic, so we must take into account the influence of electromagnetic fields. We provide the Lagrangian density for the coupling of matter field and electromagnetic field:
\begin{equation}
	\mathcal{L}=\mathcal{L}_m+\mathcal{L}_h=\mathcal{L}_m-\frac{1}{4}\tilde{F}_{\mu\nu}\tilde{F}^{\mu\nu}.
\end{equation}
The generalized coordinates corresponding to $\mathcal{L}_h$ are $\tilde{A}_\mu=(\tilde{A}_0,0,0,0)$. Obviously, $\tilde{A}_0$ is a cyclic coordinate. To eliminate the degrees of freedom corresponding to $\tilde{A}_0$, the action can be written as:
\begin{equation}
	S=\int_{t_i}^{t_f}(L-P_{\tilde{A}_0}\dot{\tilde{A}}_0)\mathrm{d}t.\label{eq:14}
\end{equation}
Due to the fact that the constant time slice of the Painlev\'{e} line element is a flat Euclidean spacetime in the radial direction, and the Reissner-Nordstr\"{o}m black hole is a steady-state black hole, the WKB approximation can be applied to it. Particle emission rate and the imaginary part of action has the following relationship:
\begin{equation}
	\Gamma\sim e^{-2\mathrm{Im}S}.\label{eq:32}
\end{equation}
According to Eq. (\ref{eq:14}), the imaginary part of the action can be written as:
\begin{equation}
	\mathrm{Im}S=\mathrm{Im}\left\{\int_{r_i}^{r_f}\left[P_r-\frac{P_{\tilde{A}_0}\dot{\tilde{A}}_0}{\dot{r}}\right]\mathrm{d}r\right\}=\mathrm{Im}\left\{\int_{r_i}^{r_f}\left[\int_{(0,0)}^{(P_r,P_{\tilde{A}_0})}\mathrm{d}P^\prime_r-\frac{\dot{\tilde{A}}_0}{\dot{r}}\mathrm{d}P^\prime_{\tilde{A}_0}\right]\mathrm{d}r\right\}.\label{eq:15}
\end{equation}
Where $P_r$ is the generalized momentum conjugate with $r$, and $P_{\tilde{A}_0}$ is the generalized momentum conjugate with $\tilde{A}_0$. $r_i$ is the initial position of the particle radiation process, slightly within the event horizon of the black hole. $r_f$ is the final position of the particle radiation process, slightly outside the event horizon of the black hole.

To proceed with our calculations, we substitute the Hamilton's equation into Eq. (\ref{eq:15}):
\begin{equation}
	\dot{r}=\frac{\mathrm{d}H}{\mathrm{d}P_r}\bigg|_{(r;\tilde{A}_0,P_{\tilde{A}_0})}.\label{eq:16}
\end{equation}
\begin{equation}
	\dot{\tilde{A}}_0=\frac{\mathrm{d}H}{\mathrm{d}P_{\tilde{A}_0}}\bigg|_{(\tilde{A}_0;r,P_{r})}.\label{eq:17}
\end{equation}
By substituting Eq. (\ref{eq:16}) and Eq. (\ref{eq:17}) into Eq. (\ref{eq:15}), we can obtain:
\begin{equation}
	\mathrm{Im}S=\mathrm{Im}\left\{\int_{r_i}^{r_f}\left[\int_{(M,E_{Q_h})}^{(M-\omega,E_{Q_h-q_h})}\frac{1}{\dot{r}}(\mathrm{d}H)_{r;\tilde{A}_0,P_{\tilde{A}_0}}-\frac{1}{\dot{r}}(\mathrm{d}H)_{\tilde{A}_0;r,P_{r}}\right]\mathrm{d}r\right\}.\label{eq:18}
\end{equation}
Where $E_{Q_h}$ is the energy corresponding to the electromagnetic field. $M$, $E_{Q_h}$ are a fixed and unchanging quantity. $M-\omega$, $E_{Q_h-q_h}$ are the residual gravitational and electromagnetic energy of the black hole after radiation. It can be clearly seen that in our formula, energy conservation, charge conservation, and magnetic charge conservation are all reflected. It is worth emphasizing that in order to simplify the equivalent charge of a black hole after radiation to $Q_h-q_h$, we adopt the following assumption:
\begin{equation}
	\frac{q_e}{q_g}=\frac{Q_e}{Q_g}.\label{eq:70}
\end{equation}
Where $q_e$ and $q_g$ are the charges and magnetic charges of the emitted particles, respectively. As mentioned above, $Q_h$ is equivalent charge of black hole corresponding to density $\rho_h$. And $q_h$ is equivalent charge of the emitted particles also corresponding to density $\rho_h$. So we can get Eq. (\ref{eq:70}). The loss of gravitational energy and electromagnetic energy caused by particle tunneling in a black hole is represented by the following equations:
\begin{equation}
	(\mathrm{d}H)_{r;\tilde{A}_0,P_{\tilde{A}_0}}=\mathrm{d}(M-\omega^\prime)=-\mathrm{d}\omega^\prime.\label{eq:19}
\end{equation}
\begin{equation}
	(\mathrm{d}H)_{\tilde{A}_0;r,P_{r}}=-\frac{Q_h-q_h^\prime}{r}\mathrm{d}q^\prime_h.\label{eq:20}
\end{equation}
By substituting Eq. (\ref{eq:19}) and Eq. (\ref{eq:20}) into Eq. (\ref{eq:18}), we can obtain:
\begin{equation}
	\mathrm{Im}S=-\mathrm{Im}\left\{\int_{r_i}^{r_f}\left[\int_{(0,0)}^{(\omega,q_h)}\frac{1}{\dot{r}}\left(\mathrm{d}\omega^\prime-\frac{Q_h-q_h^\prime}{r}\mathrm{d}q^\prime_h\right)\right]\mathrm{d}r\right\}.\label{eq:21}
\end{equation}
For the process of particles tunneling out of the event horizon of a black hole, $\dot{r}$ takes a positive sign, and accounting for the self-gravitational force, we make the following substitution: $M\to M-\omega^\prime$, $Q_h\to Q_h-q^\prime_h$, $r_{non}\to r_{non}^\prime$, $r_-\to r_-^\prime$, $r_+\to r_+^\prime$, $r_c\to r_c^\prime$, that is:
\begin{equation}
	\dot{r}=\frac{1}{2r}\sqrt{\frac{\Lambda}{3}}\frac{(r-r_{non}^\prime)(r-r_-^\prime)(r-r_+^\prime)(r-r_c^\prime)}{\sqrt{r^4+\frac{6(M-\omega^\prime)}{\Lambda}r-\frac{3}{\Lambda}(Q_h-q_h^\prime)^2}}.\label{eq:22}
\end{equation}
Substituting Eq. (\ref{eq:22}) into Eq. (\ref{eq:21}), we have:
\begin{equation}
	\mathrm{Im}S=-\mathrm{Im}\left\{\int_{r_i}^{r_f}\left[\int_{(0,0)}^{(\omega,q_h)}\frac{2r\sqrt{r^4+\frac{6(M-\omega^\prime)}{\Lambda}r-\frac{3}{\Lambda}(Q_h-q_h^\prime)^2}}{\sqrt{\frac{\Lambda}{3}}(r-r_{non}^\prime)(r-r_-^\prime)(r-r_+^\prime)(r-r_c^\prime)}\left(\mathrm{d}\omega^\prime-\frac{Q_h-q_h^\prime}{r}\mathrm{d}q^\prime_h\right)\right]\mathrm{d}r\right\}.
\end{equation}
To proceed with the calculation, switch the order of integration and integrate $r$ first. It can be clearly observed that $r=r_+^\prime$ is a pole. By selecting a new integral path to bypass the poles and applying the residue theorem, we can obtain:
\begin{equation}
	\mathrm{Im}S=-\frac{6\pi}{\Lambda}\int_{(0,0)}^{(\omega,q_h)}\frac{r_+^{\prime 2}}{(r_+^\prime-r_{non}^\prime)(r_+^\prime-r_-^\prime)(r_+^\prime-r_c^\prime)} \left(\mathrm{d}\omega^\prime-\frac{Q_h-q_h^\prime}{r_+^\prime}\mathrm{d}q^\prime_h\right).\label{eq:23}
\end{equation}

According to the relationship between $r_{non}^\prime$, $r_-^\prime$, $r_+^\prime$, $r_c^\prime$ and $\omega^\prime$, $q_h^\prime$, we can obtain:
\begin{equation}
	\frac{6}{\Lambda}[r_+^\prime\mathrm{d}\omega^\prime-(Q_h-q_h^\prime)\mathrm{d}q^\prime_h]=(r_+^\prime-r_{non}^\prime)(r_+^\prime-r_-^\prime)(r_+^\prime-r_c^\prime)\mathrm{d}r_+^\prime.
\end{equation}
Continuing the integration operation, through simple calculations, we can obtain the final result as:
\begin{equation}
	\mathrm{Im}S=-\pi\int_{r_i}^{r_f}r_+^\prime\mathrm{d}r_+^\prime=-\frac{\pi}{2}(r_f^2-r_i^2)=-\frac{1}{2}\Delta S_{BH}.
\end{equation}
Where $\Delta S_{BH}=S_{BH}(M-\omega,Q_e-q_e,Q_g-q_g)-S_{BH}(M,Q_e,Q_g)$ is the change in entropy before and after the black hole radiates particles. So the emission rate of particles can be expressed by the following equation:
\begin{equation}
	\Gamma\sim e^{-2\mathrm{Im}S}=e^{\Delta S_{BH}}.
\end{equation}
This result conforms to the unitary principle and supports the conservation of information.
\subsection{Tunneling into the cosmic horizon}
Next, we will discuss the particle tunneling process near the de Sitter cosmic horizon. Obviously, for the process of particles tunneling from outside the cosmic horizon to inside the cosmic horizon, the final position is smaller than the initial position, $r_f^\prime<r_i^\prime$. The total energy, total charge, and total magnetic charge of space-time will all increase, $M\to M+\omega^\prime$, $Q_e\to Q_e+q^\prime_e$, $Q_h\to Q_h+q^\prime_h$, $Q_h\to Q_h+q^\prime_h$. And the position of the horizon will also change, $r_{non}\to r_{non}^{\prime\prime}$, $r_-\to r_-^{\prime\prime}$, $r_+\to r_+^{\prime\prime}$, $r_c\to r_c^{\prime\prime}$. What we are considering now is the tunneling of the cosmic horizon, so $\dot{r}$ should take the negative sign. Considering the influence of self gravity, the geodesic equation becomes:
\begin{equation}
	\dot{r}=-\frac{1}{2r}\sqrt{\frac{\Lambda}{3}}\frac{(r-r_{non}^{\prime\prime})(r-r_-^{\prime\prime})(r-r_+^{\prime\prime})(r-r_c^{\prime\prime})}{\sqrt{r^4+\frac{6(M+\omega^\prime)}{\Lambda}r-\frac{3}{\Lambda}(Q_h+q_h^\prime)^2}}.\label{eq:43}
\end{equation}
Similarly, we consider the incident particle as a de Broglie s-wave and subtract the degrees of freedom of the cyclic coordinate $\tilde{A}_0$ to obtain the imaginary part of the action as:
\begin{equation}
	\mathrm{Im}S=\mathrm{Im}\left\{\int_{r_i^\prime}^{r_f^\prime}\left[P_r-\frac{P_{\tilde{A}_0}\dot{\tilde{A}}_0}{\dot{r}}\right]\mathrm{d}r\right\}=\mathrm{Im}\left\{\int_{r_i^\prime}^{r_f^\prime}\left[\int_{(0,0)}^{(P_r,P_{\tilde{A}_0})}\mathrm{d}P^\prime_r-\frac{\dot{\tilde{A}}_0}{\dot{r}}\mathrm{d}P^\prime_{\tilde{A}_t}\right]\mathrm{d}r\right\}.\label{eq:24}
\end{equation}
$r_i^\prime$ is the initial position of particle radiation, slightly outside the cosmic horizon. And $r_f^\prime$ is the final position of particle radiation, slightly within the cosmic horizon.

The Hamilton's equation is:
\begin{equation}
	\dot{r}=\frac{\mathrm{d}H}{\mathrm{d}P_r}\bigg|_{(r;\tilde{A}_0,P_{\tilde{A}_0})}=\frac{\mathrm{d}(M+\omega^\prime)}{\mathrm{d}P_r}=\frac{\mathrm{d}\omega^\prime}{\mathrm{d}P_r}.\label{eq:25}
\end{equation}
\begin{equation}
	\dot{\tilde{A}}_0=\frac{\mathrm{d}H}{\mathrm{d}P_{\tilde{A}_0}}\bigg|_{(\tilde{A}_0;r,P_{r})}=\frac{Q_h+q_h^\prime}{r}\frac{\mathrm{d}q^\prime_h}{\mathrm{d}P_{\tilde{A}_0}}.\label{eq:26}
\end{equation}
Eq. (\ref{eq:25}) and Eq. (\ref{eq:26}) respectively represent the changes in gravitational energy and electromagnetic energy after particles enter the cosmic horizon, from which it can be clearly seen that energy conservation, charge conservation, and magnetic charge conservation occur. Substituting Eq. (\ref{eq:23}), Eq. (\ref{eq:25}) and Eq. (\ref{eq:26}) into Eq. (\ref{eq:24}) yields:
\begin{equation}
	\mathrm{Im}S=-\mathrm{Im}\left\{\int_{r_i^\prime}^{r_f^\prime}\left[\int_{(0,0)}^{(\omega,q_h)}\frac{2r\sqrt{r^4+\frac{6(M+\omega^\prime)}{\Lambda}r-\frac{3}{\Lambda}(Q_h+q_h^\prime)^2}}{\sqrt{\frac{\Lambda}{3}}(r-r_{non}^{\prime\prime})(r-r_-^{\prime\prime})(r-r_+^{\prime\prime})(r-r_c^{\prime\prime})}\left(\mathrm{d}\omega^\prime-\frac{Q_h+q_h^\prime}{r}\mathrm{d}q^\prime_h\right)\right]\mathrm{d}r\right\}.
\end{equation}
Obviously, there is a pole at the cosmic horizon $r=r_+^{\prime\prime}$. In order to proceed with the calculation, exchange the order of integration, first integrate $r$ and apply the residue theorem, which yields:
\begin{equation}
	\mathrm{Im}S=-\frac{6\pi}{\Lambda}\int_{(0,0)}^{(\omega,q_h)}\frac{r_c^{\prime\prime 2}}{(r_c^{\prime\prime}-r_{non}^{\prime\prime})(r_c^{\prime\prime}-r_-^{\prime\prime})(r_c^{\prime\prime}-r_+^{\prime\prime})} \left(\mathrm{d}\omega^\prime-\frac{Q_h+q_h^\prime}{r_c^{\prime\prime}}\mathrm{d}q^\prime_h\right).\label{eq:27}
\end{equation}

According to the relationship between $r_{non}^{\prime\prime}$, $r_-^{\prime\prime}$, $r_+^{\prime\prime}$, $r_c^{\prime\prime}$ and $\omega^\prime$, $q_h^\prime$, we can obtain:
\begin{equation}
	\frac{6}{\Lambda}[r_c^{\prime\prime}\mathrm{d}\omega^\prime-(Q_h+q_h^\prime)\mathrm{d}q^\prime_h]=(r_c^{\prime\prime}-r_{non}^{\prime\prime})(r_c^{\prime\prime}-r_-^{\prime\prime})(r_c^{\prime\prime}-r_+^{\prime\prime})\mathrm{d}r_c^{\prime\prime}.
\end{equation}
By completing the integration operation through simple calculations, the final result obtained is:
\begin{equation}
	\mathrm{Im}S=-\pi\int_{r_i^\prime}^{r_f^\prime}r_c^{\prime\prime}\mathrm{d}r_c^{\prime\prime}=-\frac{\pi}{2}(r_f^{\prime 2}-r_i^{\prime 2})=-\frac{1}{2}\Delta S_{CH}.
\end{equation}
Where $\Delta S_{CH}=S_{CH}(M+\omega,Q_e+q_e,Q_g+q_g)-S_{CH}(M,Q_e,Q_g)$ is the change in entropy of particles before and after passing through the cosmic horizon. The emission rate of particles can be expressed by the following equation:
\begin{equation}
	\Gamma\sim e^{-2\mathrm{Im}S}=e^{\Delta S_{CH}}.
\end{equation}
Obviously, such a result follows the unitary principle and supports the conservation of information.
\subsection{Discussion}
The above results can be easily obtained from the first law of black hole thermodynamics, and we can obtain the temperature at the surface of the black hole event horizon and the cosmic horizon from it. For the case of magnetic charge, the first law of black hole thermodynamics should be written as:
\begin{equation}
	\mathrm{d}S^\prime=\frac{1}{T^\prime}\mathrm{d}M^\prime-\frac{V_e^\prime}{T^\prime}\mathrm{d}Q_e^\prime-\frac{V_g^\prime}{T^\prime}\mathrm{d}Q_g^\prime=\frac{1}{T^\prime}(\mathrm{d}M^\prime-V_h^\prime\mathrm{d}Q_h^\prime).
\end{equation}
Among them, for the process of particles passing through the event horizon of a black hole, $\mathrm{d}M^\prime=-\mathrm{d}\omega^\prime$, $\mathrm{d}Q_h^\prime=-\mathrm{d}q_h^\prime$, $V_h^\prime=\frac{Q_h-q_h^\prime}{r_+^\prime}$. For the process of particles crossing the cosmic horizon, $\mathrm{d}M^\prime=\mathrm{d}\omega^\prime$, $\mathrm{d}Q_h^\prime=\mathrm{d}q_h^\prime$, $V_h^\prime=\frac{Q_h+q_h^\prime}{r_c^{\prime\prime}}$. So:
\begin{equation}
	\mathrm{d}S_{BH}^\prime=-\frac{1}{T^\prime}\left(\mathrm{d}\omega^\prime-\frac{Q_h-q_h^\prime}{r_+^\prime}\mathrm{d}q_h^\prime\right).\label{eq:28}
\end{equation}
\begin{equation}
	\mathrm{d}S_{CH}^\prime=\frac{1}{T^{\prime\prime}}\left(\mathrm{d}\omega^\prime-\frac{Q_h+q_h^\prime}{r_c^{\prime\prime}}\mathrm{d}q_h^\prime\right).\label{eq:29}
\end{equation}
Comparing Eq. (\ref{eq:28}) and Eq. (\ref{eq:29}) with Eq. (\ref{eq:23}) and Eq. (\ref{eq:27}) respectively, the temperature $T^\prime$ at the event horizon surface of the black hole and the temperature $T^{\prime\prime}$ at the cosmic horizon surface are obtained:
\begin{equation}
	\beta^\prime=\frac{1}{T^\prime}=-\frac{12\pi}{\Lambda}\frac{r_+^{\prime 2}}{(r_+^\prime-r_{non}^\prime)(r_+^\prime-r_-^\prime)(r_+^\prime-r_c^\prime)}=\lim_{r\to r_+^\prime}\frac{4\pi}{|f^\prime(r)|}.
\end{equation}
\begin{equation}
	\beta^{\prime\prime}=\frac{1}{T^{\prime\prime}}=\frac{12\pi}{\Lambda}\frac{r_c^{\prime\prime 2}}{(r_c^{\prime\prime}-r_{non}^{\prime\prime})(r_c^{\prime\prime}-r_-^{\prime\prime})(r_c^{\prime\prime}-r_+^{\prime\prime})}=\lim_{r\to r_c^{\prime\prime}}\frac{4\pi}{|f^\prime(r)|}.
\end{equation}
Where $f(r)=-\frac{\Lambda}{3r^2}(r-r_{non})(r-r_-)(r-r_+)(r-r_c)$. This is consistent with the conclusion drawn from general black hole thermodynamics.
\section{Kerr-Newman-Kasuya de Sitter black hole}
\label{sec:2}
The line element of Kerr Newman Kasuya de Sitter black hole is:
\begin{equation}
	\mathrm{d}s^2=g_{00}\mathrm{d}t_k^2+g_{11}\mathrm{d}r^2+g_{22}\mathrm{d}\theta^2+g_{33}\mathrm{d}\phi^2+2g_{03}\mathrm{d}t_k\mathrm{d}\phi.
\end{equation}
Where:
\begin{align}
	g_{00}=&-\frac{\Delta_r-a^2\Delta_\theta\sin^2\theta}{\Xi^2\Sigma},\\
	g_{11}=&\frac{\Sigma}{\Delta_r},\\
	g_{22}=&\frac{\Sigma}{\Delta_\theta},\\
	g_{33}=&-\frac{h^2\Delta_r-\xi^2\Delta_\theta\sin^2\theta}{\Xi^2\Sigma},\\
	g_{03}=&\frac{h\Delta_r-a\xi\Delta_\theta\sin^2\theta}{\Xi^2\Sigma}.
\end{align}
Where $\Xi=1+\frac{\Lambda}{3}a^2$, $h=a\sin^2\theta$, $\Sigma=r^2+a^2\cos^2\theta$, $\xi=a^2+r^2$, $\Delta_\theta=1+\frac{\Lambda}{3}a^2\cos^2\theta$, $\Delta_r=\xi\left(1-\frac{\Lambda}{3}r^2\right)-2Mr+Q_h^2$, $Q_h^2=Q_e^2+Q_g^2$. $M$ is the mass of the black hole. $Q_e$ and $Q_g$ are the charges and magnetic charges of the black hole, respectively. $a=\frac{J}{M}$ is the angular momentum per unit mass of a black hole.

The position of the horizon are given by equation $\Delta_r=0$, that is:
\begin{equation}
	\Delta_r=-\frac{\Lambda}{3}r^4+\left(1-\frac{\Lambda}{3}a^2\right)r^2-2Mr+Q_h^2+a^2=0.\label{eq:30}
\end{equation}
Let $A=\frac{3}{\Lambda}(a^2+Q_h^2)$, $B=\frac{3}{\Lambda}-a^2$. The equation Eq. \eqref{eq:30} has four solutions, $r_{non}$, $r_-$, $r_+$, $r_c$. Among them, $r_{non}$ is a negative value solution with no physical meaning, $r_-$ is the inner event horizon of black hole, $r_+$ is the outer event horizon of black hole, and $r_c$ is the cosmic horizon. The specific forms of these four solutions are:
\begin{align}
	r_{non}=&-\frac{1}{2}\tilde\eta-\frac{1}{2}\tilde\zeta,\\
	r_-=&-\frac{1}{2}\tilde\eta+\frac{1}{2}\tilde\zeta,\\
	r_+=&\frac{1}{2}\tilde\eta-\frac{1}{2}\tilde\zeta,\\
	r_c=&\frac{1}{2}\tilde\eta+\frac{1}{2}\tilde\zeta.
\end{align}
Where:
\begin{align}
	\tilde\eta=&\sqrt{\frac{2B}{3}+\frac{(2)^{1/3}(\Lambda^2B^2-12\Lambda^2A)}{3\Lambda\tilde\lambda}+\frac{\tilde\lambda}{3(2)^{1/3}\Lambda}},\\
	\tilde\zeta=&\sqrt{\frac{4B}{3}-\frac{(2)^{1/3}(\Lambda^2B^2-12\Lambda^2A)}{3\Lambda\tilde\lambda}-\frac{\tilde\lambda}{3(2)^{1/3}\Lambda}-\frac{12M}{\Lambda\tilde\eta}},\\
	\tilde\lambda=&\left[972\Lambda M^2-72\Lambda^3AB-2\Lambda^3B^3+\right.\nonumber\\
	&\left.\sqrt{(72\Lambda^3AB+2\Lambda^3B^3-972\Lambda M^2)^2-4(\Lambda^2B^2-12\Lambda^2A)^3}\right]^{1/3}.
\end{align}
For the process of particles tunneling out of the event horizon of a black hole, the initial position of the particle is:
\begin{equation}
	r_i=\frac{1}{2}\tilde\eta-\frac{1}{2}\tilde\zeta.
\end{equation}
The final position of the particle is:
\begin{equation}
	r_f=\frac{1}{2}\tilde\eta_f-\frac{1}{2}\tilde\zeta_f.
\end{equation}
Where:
\begin{align}
	\tilde\eta_f=&\sqrt{\frac{2B_f}{3}+\frac{(2)^{1/3}(\Lambda^2B_f^2-12\Lambda^2A_f)}{3\Lambda\tilde\lambda_f}+\frac{\tilde\lambda_f}{3(2)^{1/3}\Lambda}},\\
	\tilde\zeta_f=&\sqrt{\frac{4B_f}{3}-\frac{(2)^{1/3}(\Lambda^2B_f^2-12\Lambda^2A_f)}{3\Lambda\tilde\lambda_f}-\frac{\tilde\lambda_f}{3(2)^{1/3}\Lambda}-\frac{12(M-\omega)}{\Lambda\tilde\eta_f}},\\
	\tilde\lambda_f=&\left\{972\Lambda (M-\omega)^2-72\Lambda^3A_fB_f-2\Lambda^3B_f^3+\right.\nonumber\\
	&\left.\sqrt{[72\Lambda^3A_fB_f+2\Lambda^3B_f^3-972\Lambda (M-\omega)^2]^2-4(\Lambda^2B_f^2-12\Lambda^2A_f)^3}\right\}^{1/3}.
\end{align}
Where $A_f=\frac{3}{\Lambda}[a_f^2+(Q_h-q_h)^2]$, $B_f=\frac{3}{\Lambda}-a_f^2$, $a_f=\frac{J-j}{M-\omega}$. After the particle exits, the position of horizons change to:
\begin{align}
	r_{non}^\prime=&-\frac{1}{2}\tilde\eta^\prime-\frac{1}{2}\tilde\zeta^\prime,\\
	r_-^\prime=&-\frac{1}{2}\tilde\eta^\prime+\frac{1}{2}\tilde\zeta^\prime,\\
	r_+^\prime=&\frac{1}{2}\tilde\eta^\prime-\frac{1}{2}\tilde\zeta^\prime,\\
	r_c^\prime=&\frac{1}{2}\tilde\eta^\prime+\frac{1}{2}\tilde\zeta^\prime.
\end{align}
Where:
\begin{align}
	\tilde\eta^\prime=&\sqrt{\frac{2B^\prime}{3}+\frac{(2)^{1/3}(\Lambda^2B^{\prime 2}-12\Lambda^2A^\prime)}{3\Lambda\tilde\lambda^\prime}+\frac{\tilde\lambda^\prime}{3(2)^{1/3}\Lambda}},\\
	\tilde\zeta^\prime=&\sqrt{\frac{4B^\prime}{3}-\frac{(2)^{1/3}(\Lambda^2B^{\prime 2}-12\Lambda^2A^\prime)}{3\Lambda\tilde\lambda^\prime}-\frac{\tilde\lambda^\prime}{3(2)^{1/3}\Lambda}-\frac{12(M-\omega^\prime)}{\Lambda\tilde\eta^\prime}},\\
	\tilde\lambda^\prime=&\left\{972\Lambda (M-\omega^\prime)^2-72\Lambda^3A^\prime B^\prime-2\Lambda^3B^{\prime 3}+\right.\nonumber\\
	&\left.\sqrt{[72\Lambda^3A^\prime B^\prime+2\Lambda^3B^{\prime 3}-972\Lambda (M-\omega^\prime)^2]^2-4(\Lambda^2B^{\prime 2}-12\Lambda^2A^\prime)^3}\right\}^{1/3}.
\end{align}
Where $A^\prime=\frac{3}{\Lambda}[a^{\prime 2}+(Q_h-q_h^\prime)^2]$, $B^\prime=\frac{3}{\Lambda}-a^{\prime 2}$, $a^\prime=\frac{J-j^\prime}{M-\omega^\prime}$.

For the process of particle tunneling into the cosmic horizon, the initial position of the particle is:
\begin{equation}
	r_i^\prime=\frac{1}{2}\tilde\eta+\frac{1}{2}\tilde\zeta.
\end{equation}
The final position of the particle is:
\begin{equation}
	r_f^\prime=\frac{1}{2}\tilde\eta_f^\prime+\frac{1}{2}\tilde\zeta_f^\prime.
\end{equation}
Where:
\begin{align}
	\tilde\eta_f^\prime=&\sqrt{\frac{2B_f^\prime}{3}+\frac{(2)^{1/3}(\Lambda^2B_f^{\prime 2}-12\Lambda^2A_f^\prime)}{3\Lambda\tilde\lambda_f^\prime}+\frac{\tilde\lambda_f^\prime}{3(2)^{1/3}\Lambda}},\\
	\tilde\zeta_f^\prime=&\sqrt{\frac{4B_f^\prime}{3}-\frac{(2)^{1/3}(\Lambda^2B_f^{\prime 2}-12\Lambda^2A_f^\prime)}{3\Lambda\tilde\lambda_f^\prime}-\frac{\tilde\lambda_f^\prime}{3(2)^{1/3}\Lambda}-\frac{12(M+\omega)}{\Lambda\tilde\eta_f^\prime}},\\
	\tilde\lambda_f^\prime=&\left\{972\Lambda (M+\omega)^2-72\Lambda^3A_f^\prime B_f^\prime-2\Lambda^3B_f^{\prime 3}+\right.\nonumber\\
	&\left.\sqrt{[72\Lambda^3A_f^\prime B_f^\prime+2\Lambda^3B_f^{\prime 3}-972\Lambda (M+\omega)^2]^2-4(\Lambda^2B_f^{\prime 2}-12\Lambda^2A_f^\prime)^3}\right\}^{1/3}.
\end{align}
Where $A_f^\prime=\frac{3}{\Lambda}[a_f^{\prime 2}+(Q_h+q_h)^2]$, $B_f^\prime=\frac{3}{\Lambda}-a_f^{\prime 2}$, $a_f^\prime=\frac{J+j}{M+\omega}$. After the particle is incident, the position of horizons change to:
\begin{align}
	r_{non}^{\prime\prime}=&-\frac{1}{2}\tilde\eta^{\prime\prime}-\frac{1}{2}\tilde\zeta^{\prime\prime},\\
	r_-^{\prime\prime}=&-\frac{1}{2}\tilde\eta^{\prime\prime}+\frac{1}{2}\tilde\zeta^{\prime\prime},\\
	r_+^{\prime\prime}=&\frac{1}{2}\tilde\eta^{\prime\prime}-\frac{1}{2}\tilde\zeta^{\prime\prime},\\
	r_c^{\prime\prime}=&\frac{1}{2}\tilde\eta^{\prime\prime}+\frac{1}{2}\tilde\zeta^{\prime\prime}.
\end{align}
Where:
\begin{align}
	\tilde\eta^{\prime\prime}=&\sqrt{\frac{2B^{\prime\prime}}{3}+\frac{(2)^{1/3}(\Lambda^2B^{\prime\prime 2}-12\Lambda^2A^{\prime\prime})}{3\Lambda\tilde\lambda^{\prime\prime}}+\frac{\tilde\lambda^{\prime\prime}}{3(2)^{1/3}\Lambda}},\\
	\tilde\zeta^{\prime\prime}=&\sqrt{\frac{4B^{\prime\prime}}{3}-\frac{(2)^{1/3}(\Lambda^2B^{\prime\prime 2}-12\Lambda^2A^{\prime\prime})}{3\Lambda\tilde\lambda^{\prime\prime}}-\frac{\tilde\lambda^{\prime\prime}}{3(2)^{1/3}\Lambda}-\frac{12(M+\omega^\prime)}{\Lambda\tilde\eta^{\prime\prime}}},\\
	\tilde\lambda^{\prime\prime}=&\left\{972\Lambda (M+\omega^\prime)^2-72\Lambda^3A^{\prime\prime} B^{\prime\prime}-2\Lambda^3B^{\prime\prime 3}+\right.\nonumber\\
	&\left.\sqrt{[72\Lambda^3A^{\prime\prime}B^{\prime\prime}+2\Lambda^3B^{\prime\prime 3}-972\Lambda (M+\omega^\prime)^2]^2-4(\Lambda^2B^{\prime\prime 2}-12\Lambda^2A^{\prime\prime})^3}\right\}^{1/3}.
\end{align}
Where $A^{\prime\prime}=\frac{3}{\Lambda}[a^{\prime\prime 2}+(Q_h+q_h^\prime)^2]$, $B^{\prime\prime}=\frac{3}{\Lambda}-a^{\prime\prime 2}$, $a^{\prime\prime}=\frac{J+j^\prime}{M+\omega^\prime}$.
\subsection{Painlev\'{e} coordinate and time-like geodesic line equation}
To eliminate coordinate singularity at the horizon, we first introduce a dragged coordinate system for the Kerr Newman Kasuya de Sitter spacetime. Let:
\begin{equation}
	\Omega=\frac{\mathrm{d}\phi}{\mathrm{d}t_k}=-\frac{g_{03}}{g_{33}}=\frac{h\Delta_r-a\xi\Delta_\theta\sin^2\theta}{h^2\Delta_r-\xi^2\Delta_\theta\sin^2\theta}.
\end{equation}
\begin{equation}
	\hat{g}_{00}=g_{00}-\frac{g_{03}^2}{g_{33}}=\frac{\Sigma\Delta_\theta\sin^2\theta\Delta_r}{\Xi^2(h^2\Delta_r-\xi^2\Delta_\theta\sin^2\theta)}.
\end{equation}
The line elements of Kerr Newman Kasuya de Sitter black hole can be rewritten as:
\begin{equation}
	\mathrm{d}s^2=\hat{g}_{00}\mathrm{d}t_k^2+g_{11}\mathrm{d}r^2+g_{22}\mathrm{d}\theta^2=\frac{\Sigma\Delta_\theta\sin^2\theta\Delta_r}{\Xi^2(h^2\Delta_r-\xi^2\Delta_\theta\sin^2\theta)}\mathrm{d}t_k^2+\frac{\Sigma}{\Delta_r}\mathrm{d}r^2+\frac{\Sigma}{\Delta_\theta}\mathrm{d}\theta^2.
\end{equation}
Due to the fact that the constant time slice of the dragging coordinate system is not a flat Euclidean space in the radial direction, it is not the coordinate system we want. We adopt a method similar to ref.~\cite{24} to perform coordinate transformation again:
\begin{equation}
	\mathrm{d}t_k=\mathrm{d}t+F(r,\theta)\mathrm{d}r+G(r,\theta)\mathrm{d}\theta.
\end{equation}
\begin{equation}
	t=t_k-\int[F(r,\theta)\mathrm{d}r+G(r,\theta)\mathrm{d}\theta].
\end{equation}
$F(r,\theta)$ and $G(r,\theta)$ satisfy the following relationship:
\begin{equation}
	\frac{\partial F(r,\theta)}{\partial\theta}=\frac{\partial G(r,\theta)}{\partial r}.
\end{equation}
In order to make the constant time slice of the new coordinates a flat Euclidean space in the radial direction, we require $F(r,\theta)$ to satisfy the following relationship:
\begin{equation}
	g_{11}+\hat{g}_{00}F^2(r,\theta)=1.
\end{equation}
So we obtained the coordinates of Painlev\'{e}-Kerr-Newman-Kasuya de Sitter:
\begin{align}
	\mathrm{d}s^2=&\hat{g}_{00}\mathrm{d}t^2\pm 2\sqrt{\hat{g}_{00}(1-g_{11})}\mathrm{d}t\mathrm{d}r+\mathrm{d}r^2+[\hat{g}_{00}G^2(r,\theta)+g_{22}]\mathrm{d}\theta^2\nonumber\\
	&+2\hat{g}_{00}G(r,\theta)\mathrm{d}t\mathrm{d}\theta+2\sqrt{\hat{g}_{00}(1-g_{11})}G(r,\theta)\mathrm{d}r\mathrm{d}\theta.
\end{align}

Similar to Section \ref{sec:1}, we treat particles as de Broglie s-waves and derive the time-like geodesic equation for particle motion using the WKB approximate:
\begin{equation}
	\dot{r}=-\frac{1}{2}\frac{\hat{g}_{00}}{\hat{g}_{01}}=\pm\frac{\Sigma\Delta_r}{2\Xi}\sqrt{\frac{\Delta_\theta\sin^2\theta}{(\Delta_r-\Sigma)\Sigma(h^2\Delta_r-\xi^2\Delta_\theta\sin^2\theta)}}.\label{eq:31}
\end{equation}
Where $\hat{g}_{01}=\sqrt{\hat{g}_{00}(1-g_{11})}$. When a particle exits the event horizon of a black hole, Eq. (\ref{eq:31}) takes a positive sign. When a particle is incident from the cosmic horizon, Eq. \eqref{eq:31} takes a negative sign. When considering self gravity, replace Eq. (\ref{eq:31}) with the following, $M\to M\mp\omega$, $Q_h\to Q_h-q_h$, $a=\frac{J}{M}\to\frac{J\mp j}{M\mp\omega}=\tilde{a}$. The minus sign and plus sign represent the outgoing particles and incoming particles, respectively.
\subsection{Tunneling out of the black hole outer event horizon}
Similar to Section \ref{sec:1}, it can be inferred from the Lagrangian density of matter-electromagnetic field that:
\begin{equation}
	\mathcal{L}=\mathcal{L}_m+\mathcal{L}_h=\mathcal{L}_m-\frac{1}{4}\tilde{F}_{\mu\nu}\tilde{F}^{\mu\nu}.
\end{equation}
$\tilde{A}_0$ is the cyclic coordinate. Under the condition of rotating with electricity, $\tilde{A}_0=-\frac{Q_hr}{\Sigma}(1-a\Omega\sin^2\theta)$. The electromagnetic potential at the event horizon of a black hole is $\tilde{A}_0|_{r_+}=-\frac{Q_hr_+}{r_+^2+a^2}$ and it at the cosmic horizon is $\tilde{A}_0|_{r_c}=-\frac{Q_hr_c}{r_c^2+a^2}$. When the line element of a rotating black hole is written in a dragged coordinate system, the line element does not contain $\phi$, so $\phi$ is also a cyclic coordinate. To eliminate the degrees of freedom corresponding to $\phi$ and $\tilde{A}_0$, the action can be written as:
\begin{equation}
	S=\int^{t_f}_{t_i}(L-P_{\tilde{A}_0}\dot{\tilde{A}}_0-P_\phi\dot{\phi})\mathrm{d}t.
\end{equation}
For the same reason as in Section \ref{sec:1}, the WKB approximation can be applied. The emission rate of particles has the same relationship as Eq. (\ref{eq:32}):
\begin{equation}
	\Gamma\sim e^{-2\mathrm{Im}S}.
\end{equation}
We obtain the imaginary part of the action as:
\begin{align}
	\mathrm{Im}S=&\mathrm{Im}\left\{\int_{r_i}^{r_f}\left[P_r-\frac{P_{\tilde{A}_0}\dot{\tilde{A}}_0}{\dot{r}}-\frac{P_\phi\dot{\phi}}{\dot{r}}\right]\mathrm{d}r\right\}\nonumber\\
	=&\mathrm{Im}\left\{\int_{r_i}^{r_f}\left[\int_{(0,0)}^{(P_r,P_{\tilde{A}_0})}\mathrm{d}P^\prime_r-\frac{\dot{\tilde{A}}_0}{\dot{r}}\mathrm{d}P^\prime_{\tilde{A}_0}-\frac{\dot{\phi}}{\dot{r}}\mathrm{d}P^\prime_{\phi}\right]\mathrm{d}r\right\}.\label{eq:33}
\end{align}

To proceed with the calculation, we write down the Hamilton's equation:
\begin{equation}
	\dot{r}=\frac{\mathrm{d}H}{\mathrm{d}P_r}\bigg|_{(r;\tilde{A}_0,P_{\tilde{A}_0};\phi,P_\phi)}=\frac{\mathrm{d}(M-\omega^\prime)}{\mathrm{d}P_r}=-\frac{\mathrm{d}\omega^\prime}{\mathrm{d}P_r}.\label{eq:34}
\end{equation}
\begin{equation}
	\dot{\tilde{A}}_0=\frac{\mathrm{d}H}{\mathrm{d}P_{\tilde{A}_0}}\bigg|_{(\tilde{A}_0;r,P_{r};\phi,P_\phi)}=-\frac{(Q_h-q_h^\prime)r}{r^2+a^{\prime 2}}\frac{\mathrm{d}q_h^\prime}{\mathrm{d}P_{\tilde{A}_0}}.\label{eq:35}
\end{equation}
Additionally, we have the following relationship:
\begin{equation}
	\dot{\phi}=\frac{\mathrm{d}\phi}{\mathrm{d}t}=\Omega^\prime.\label{eq:36}
\end{equation}
\begin{equation}
	P_\phi^\prime=J^\prime=J-j^\prime=(M-\omega^\prime)a^\prime.\label{eq:37}
\end{equation}
In Eq. (\ref{eq:34}), Eq. (\ref{eq:35}) and Eq. (\ref{eq:37}), conservation of energy, conservation of angular momentum, conservation of electric charge and conservation of magnetic charge are manifested. Substitute Eq. (\ref{eq:34}), Eq. (\ref{eq:35}), Eq. (\ref{eq:36}), Eq. (\ref{eq:37}) into Eq. (\ref{eq:33}), we can obtain:
\begin{equation}
	\mathrm{Im}S=-\mathrm{Im}\left\{\int_{r_i}^{r_f}\left[\int_{(0,0,0)}^{(\omega,q_h,j)}\frac{1}{\dot{r}}\left(\mathrm{d}\omega^\prime-\frac{(Q_h-q_h^\prime)r}{r^2+a^{\prime 2}}\mathrm{d}q^\prime_h-\Omega^\prime\mathrm{d}j^\prime\right)\right]\mathrm{d}r\right\}.\label{eq:38}
\end{equation}
When particles tunnel out of the event horizon of a black hole, $\dot{r}$ takes a positive sign. And considering the self gravity effect, make the following substitution, $M\to M-\omega^\prime$, $Q_h=Q_h-q_h^\prime$, $a=\frac{J}{M}\to\frac{J-j^\prime}{M-\omega^\prime}=a^\prime$, $\Xi=1+\frac{\Lambda}{3}a^2\to 1+\frac{\Lambda}{3}a^{\prime 2}=\Xi^\prime$, $h=a\sin^2\theta\to a^\prime\sin^2\theta=h^\prime$, $\Sigma=r^2+a^2\cos^2\theta\to r^2+a^{\prime 2}\cos^2\theta=\Sigma^\prime$, $\xi=a^2+r^2\to a^{\prime 2}+r^2=\xi^\prime$, $\Delta_\theta=1+\frac{\Lambda}{3}a^2\cos^2\theta\to 1+\frac{\Lambda}{3}a^{\prime 2}\cos^2\theta=\Delta_\theta^\prime$, $\Delta_r=\xi\left(1-\frac{\Lambda}{3}r^2\right)-2Mr+Q_h^2\to\xi^\prime\left(1-\frac{\Lambda}{3}r^2\right)-2(M-\omega^\prime)r+(Q_h-q_h^\prime)^2=\Delta_r^\prime$, $r_{non}\to r_{non}^\prime$, $r_-\to r_-^\prime$, $r_+\to r_+^\prime$, $r_c\to r_c^\prime$, that is:
\begin{equation}
	\dot{r}=\frac{\Sigma^\prime\Delta_r^\prime}{2\Xi^\prime}\sqrt{\frac{\Delta_\theta^\prime\sin^2\theta}{(\Delta_r^\prime-\Sigma^\prime)\Sigma^\prime(h^{\prime 2}\Delta_r^\prime-\xi^{\prime 2}\Delta_\theta^\prime\sin^2\theta)}}.\label{eq:39}
\end{equation}
Where $\Delta_r=-\frac{\Lambda}{3}(r-r_{non})(r-r_{-})(r-r_{+})(r-r_{c})$. Substitute Eq. (\ref{eq:39}) into Eq. (\ref{eq:38}), we can obtain:
\begin{align}
	\mathrm{Im}S=&-\mathrm{Im}\left\{\int_{r_i}^{r_f}\left[\int_{(0,0,0)}^{(\omega,q_h,j)}\frac{6\Xi^\prime}{\Lambda\Sigma^\prime}\frac{1}{(r-r_{non}^\prime)(r-r_{-}^\prime)(r-r_{+}^\prime)(r-r_{c}^\prime)}\times\right.\right.\nonumber\\
	&\left.\left.\sqrt{\frac{(\Delta_r^\prime-\Sigma^\prime)\Sigma^\prime(h^{\prime 2}\Delta_r^\prime-\xi^{\prime 2}\Delta_\theta^\prime\sin^2\theta)}{\Delta_\theta^\prime\sin^2\theta}}\left(\mathrm{d}\omega^\prime-\frac{(Q_h-q_h^\prime)r}{r^2+a^{\prime 2}}\mathrm{d}q^\prime_h-\Omega^\prime\mathrm{d}j^\prime\right)\right]\mathrm{d}r\right\}.
\end{align}
Switch the order of integration, integrating $r$ first. It is obvious that $r=r_+^\prime$ is a pole. It can be obtained by re-selecting the integral circumference and applying the residue theorem:
\begin{equation}
	\mathrm{Im}S=-2\pi\int_{(0,0,0)}^{(\omega,q_h,j)}\frac{(\frac{\Lambda}{3}+a^{\prime 2})(a^{\prime 2}+r_+^{\prime 2})}{(r_{+}^\prime-r_{non}^\prime)(r_{+}^\prime-r_{-}^\prime)(r_{+}^\prime-r_{c}^\prime)}\left(\mathrm{d}\omega^\prime-\frac{(Q_h-q_h^\prime)r_+^\prime}{r_+^{\prime 2}+a^{\prime 2}}\mathrm{d}q^\prime_h-\Omega^\prime\mathrm{d}j^\prime\right).\label{eq:40}
\end{equation}
Where $\Omega^\prime|_{r_+^\prime}=\frac{a^\prime}{r_+^{\prime 2}+a^{\prime 2}}$. Based on the relationship between $r_{non}^\prime$, $r_{-}^\prime$, $r_{+}^\prime$, $r_{c}^\prime$ and $\omega^\prime$, $q_h^\prime$, $j^\prime$, we can get:
\begin{equation}
	\left(\frac{6}{\Lambda}+2a^{\prime 2}\right)[(r_+^{\prime 2}+a^{\prime 2})\mathrm{d}\omega^\prime-(Q_h-q_h^\prime)r_+^\prime\mathrm{d}q^\prime_h-a^\prime\mathrm{d}j^\prime]=(r_+^\prime-r_{non}^\prime)(r_+^\prime-r_-^\prime)(r_+^\prime-r_c^\prime)\mathrm{d}r_+^\prime.\label{eq:41}
\end{equation}
By substituting Eq. (\ref{eq:41}) into Eq. (\ref{eq:40}), we get our final result:
\begin{equation}
	\mathrm{Im}S=-\pi\int_{r_i}^{r_f}r_+^\prime\mathrm{d}r_+^\prime=-\frac{\pi}{2}(r_f^2-r_i^2)=-\frac{1}{2}\Delta S_{BH}.
\end{equation}
Where $\Delta S_{BH}=S_{BH}(M-\omega,Q_e-q_e,Q_g-q_g,J-j)-S_{BH}(M,Q_e,Q_g,J)$ is the change in entropy before and after the black hole radiates particles. So the emission rate of particles can be expressed by the following equation:
\begin{equation}
	\Gamma\sim e^{-2\mathrm{Im}S}=e^{\Delta S_{BH}}.\label{eq:42}
\end{equation}
Eq. (\ref{eq:42}) indicates that our results conform to the unitary principle and support the conservation of information.
\subsection{Tunneling into the cosmic horizon}
The process of Kerr-Newman-Kasuya de Sitter space-time particles tunneling into the cosmic horizon increases the total energy, total charge, total magnetic charge, and total angular momentum of space-time, $M\to M+\omega^\prime$, $Q_h\to Q_h+q_h^\prime$, $a=\frac{J}{M}\to\frac{J+j^\prime}{M+\omega^\prime}=a^{\prime\prime}$. Other quantities that depend on $M$, $Q_h$, $J$ also change, $\Xi=1+\frac{\Lambda}{3}a^2\to 1+\frac{\Lambda}{3}a^{\prime\prime 2}=\Xi^{\prime\prime}$, $h=a\sin^2\theta\to a^{\prime\prime}\sin^2\theta=h^{\prime\prime}$, $\Sigma=r^2+a^2\cos^2\theta\to r^2+a^{\prime 2}\cos^2\theta=\Sigma^{\prime\prime}$, $\xi=a^2+r^2\to a^{\prime\prime 2}+r^2=\xi^{\prime\prime}$, $\Delta_\theta=1+\frac{\Lambda}{3}a^2\cos^2\theta\to 1+\frac{\Lambda}{3}a^{\prime\prime 2}\cos^2\theta=\Delta_\theta^{\prime\prime}$, $\Delta_r=\xi\left(1-\frac{\Lambda}{3}r^2\right)-2Mr+Q_h^2\to\xi^{\prime\prime}\left(1-\frac{\Lambda}{3}r^2\right)-2(M+\omega^\prime)r+(Q_h+q_h^\prime)^2=\Delta_r^{\prime\prime}$, $r_{non}\to r_{non}^{\prime\prime}$, $r_-\to r_-^{\prime\prime}$, $r_+\to r_+^{\prime\prime}$, $r_c\to r_c^{\prime\prime}$. For particles tunneling into the cosmic horizon, $\dot{r}$ is a negative sign. Taking into account the effect of self-gravitation, the geodesic equation becomes:
\begin{equation}
	\dot{r}=-\frac{\Sigma^{\prime\prime}\Delta_r^{\prime\prime}}{2\Xi^{\prime\prime}}\sqrt{\frac{\Delta_\theta^{\prime\prime}\sin^2\theta}{(\Delta_r^{\prime\prime}-\Sigma^{\prime\prime})\Sigma^{\prime\prime}(h^{\prime\prime 2}\Delta_r^{\prime\prime}-\xi^{\prime\prime 2}\Delta_\theta^{\prime\prime}\sin^2\theta)}}.\label{eq:52}
\end{equation}
We treat the incident particle as a de Broglie s-wave and subtract the degrees of freedom of the cyclic coordinates $\tilde{A}_0$ and $\phi$ to get the imaginary part of the action:
\begin{align}
	\mathrm{Im}S=&\mathrm{Im}\left\{\int_{r_i^\prime}^{r_f^\prime}\left[P_r-\frac{P_{\tilde{A}_0}\dot{\tilde{A}}_0}{\dot{r}}-\frac{P_\phi\dot{\phi}}{\dot{r}}\right]\mathrm{d}r\right\}\nonumber\\
	=&\mathrm{Im}\left\{\int_{r_i^\prime}^{r_f^\prime}\left[\int_{(0,0)}^{(P_r,P_{\tilde{A}_0})}\mathrm{d}P^\prime_r-\frac{\dot{\tilde{A}}_0}{\dot{r}}\mathrm{d}P^\prime_{\tilde{A}_0}-\frac{\dot{\phi}}{\dot{r}}\mathrm{d}P^\prime_{\phi}\right]\mathrm{d}r\right\}.\label{eq:44}
\end{align}
The Hamilton's equation is:
\begin{equation}
	\dot{r}=\frac{\mathrm{d}H}{\mathrm{d}P_r}\bigg|_{(r;\tilde{A}_0,P_{\tilde{A}_0};\phi,P_\phi)}=\frac{\mathrm{d}(M+\omega^\prime)}{\mathrm{d}P_r}=\frac{\mathrm{d}\omega^\prime}{\mathrm{d}P_r}.\label{eq:45}
\end{equation}
\begin{equation}
	\dot{\tilde{A}}_0=\frac{\mathrm{d}H}{\mathrm{d}P_{\tilde{A}_0}}\bigg|_{(\tilde{A}_0;r,P_{r};\phi,P_\phi)}=\frac{(Q_h+q_h^\prime)r}{r^2+a^{\prime\prime 2}}\frac{\mathrm{d}q_h^\prime}{\mathrm{d}P_{\tilde{A}_0}}.\label{eq:46}
\end{equation}
In addition, we have the following relationship:
\begin{equation}
	\dot{\phi}=\frac{\mathrm{d}\phi}{\mathrm{d}t}=\Omega^\prime.\label{eq:47}
\end{equation}
\begin{equation}
	P_\phi^\prime=J^\prime=J+j^\prime=(M+\omega^\prime)a^{\prime\prime}.\label{eq:48}
\end{equation}
Eq. (\ref{eq:45}), Eq. (\ref{eq:46}) and Eq. (\ref{eq:48}) clearly represent conservation of energy, conservation of angular momentum, conservation of charge and conservation of magnetic charge. Substitute Eq. (\ref{eq:52}), Eq. (\ref{eq:45}), Eq. (\ref{eq:46}), Eq. (\ref{eq:47}) and Eq. (\ref{eq:48}) into Eq. (\ref{eq:44}), we can obtain:
\begin{align}
	\mathrm{Im}S=&-\mathrm{Im}\left\{\int_{r_i^\prime}^{r_f^\prime}\left[\int_{(0,0,0)}^{(\omega,q_h,j)}\frac{6\Xi^{\prime\prime}}{\Lambda\Sigma^{\prime\prime}}\frac{1}{(r-r_{non}^{\prime\prime})(r-r_{-}^{\prime\prime})(r-r_{+}^{\prime\prime})(r-r_{c}^{\prime\prime})}\times\right.\right.\nonumber\\
	&\left.\left.\sqrt{\frac{(\Delta_r^{\prime\prime}-\Sigma^{\prime\prime})\Sigma^{\prime\prime}(h^{\prime\prime 2}\Delta_r^{\prime\prime}-\xi^{\prime\prime 2}\Delta_\theta^{\prime\prime}\sin^2\theta)}{\Delta_\theta^{\prime\prime}\sin^2\theta}}\left(\mathrm{d}\omega^\prime-\frac{(Q_h+q_h^\prime)r}{r^2+a^{\prime\prime 2}}\mathrm{d}q^\prime_h-\Omega^\prime\mathrm{d}j^\prime\right)\right]\mathrm{d}r\right\}.
\end{align}
Switch the order of integration, integrating $r$ first. Clearly, $r=r_{c}^{\prime\prime}$ is a pole. It can be obtained by re-selecting the integral circumference and applying the residue theorem:
\begin{equation}
	\mathrm{Im}S=-2\pi\int_{(0,0,0)}^{(\omega,q_h,j)}\frac{(\frac{\Lambda}{3}+a^{\prime\prime 2})(a^{\prime\prime 2}+r_c^{\prime\prime 2})}{(r_c^{\prime\prime}-r_{non}^{\prime\prime})(r_c^{\prime\prime}-r_-^{\prime\prime})(r_c^{\prime\prime}-r_+^{\prime\prime})}\left(\mathrm{d}\omega^\prime-\frac{(Q_h+q_h^\prime)r_c^{\prime\prime}}{r_c^{\prime\prime 2}+a^{\prime\prime 2}}\mathrm{d}q^\prime_h-\Omega^\prime\mathrm{d}j^\prime\right).\label{eq:49}
\end{equation}
Where $\Omega^\prime|_{r_c^{\prime\prime}}=\frac{a^{\prime\prime}}{r_c^{\prime\prime 2}+a^{\prime\prime 2}}$. Based on the relationship between $r_{non}^{\prime\prime}$, $r_{-}^{\prime\prime}$, $r_{+}^{\prime\prime}$, $r_{c}^{\prime\prime}$ and $\omega^\prime$, $q_h^\prime$, $j^\prime$, we can get:
\begin{equation}
	\left(\frac{6}{\Lambda}+2a^{\prime\prime 2}\right)[(r_c^{\prime\prime 2}+a^{\prime\prime 2})\mathrm{d}\omega^\prime-(Q_h+q_h^\prime)r_c^{\prime\prime}\mathrm{d}q^\prime_h-a^{\prime\prime}\mathrm{d}j^\prime]=(r_c^{\prime\prime}-r_{non}^{\prime\prime})(r_c^{\prime\prime}-r_-^{\prime\prime})(r_c^{\prime\prime}-r_+^{\prime\prime})\mathrm{d}r_c^{\prime\prime}.\label{eq:50}
\end{equation}
By substituting Eq. (\ref{eq:50}) into Eq. (\ref{eq:49}), we get our final result:
\begin{equation}
	\mathrm{Im}S=-\pi\int_{r_i^\prime}^{r_f^\prime}r_c^{\prime\prime}\mathrm{d}r_c^{\prime\prime}=-\frac{\pi}{2}(r_f^{\prime 2}-r_i^{\prime 2})=-\frac{1}{2}\Delta S_{CH}.
\end{equation}
Where $\Delta S_{CH}=S_{CH}(M+\omega,Q_e+q_e,Q_g+q_g,J+j)-S_{CH}(M,Q_e,Q_g,J)$ is the change in entropy before and after particle enters the cosmic horizon. So the emission rate of particles can be expressed by the following equation:
\begin{equation}
	\Gamma\sim e^{-2\mathrm{Im}S}=e^{\Delta S_{CH}}.\label{eq:51}
\end{equation}
Clearly, Eq. (\ref{eq:51}) indicates that our results conform to the unitary principle and support the conservation of information.
\subsection{Discussion}
Similarly, the results in this section can be easily obtained from the first law of black hole thermodynamics, and the temperature at the surface of black hole event horizon and cosmic horizon can be read from it. For the most general case of a rotating black hole that is both charged and magnetic, the first law of black hole thermodynamics can be written:
\begin{equation}
	\mathrm{d}S^\prime=\frac{1}{T^\prime}\mathrm{d}M^\prime-\frac{V_e^\prime}{T^\prime}\mathrm{d}Q_e^\prime-\frac{V_g^\prime}{T^\prime}\mathrm{d}Q_g^\prime-\frac{\Omega^\prime}{T^\prime}\mathrm{d}J^\prime=\frac{1}{T^\prime}(\mathrm{d}M^\prime-V_h^\prime\mathrm{d}Q_h^\prime-\Omega^\prime\mathrm{d}J^\prime).
\end{equation}
For the process by which particles tunnel out of a black hole's event horizon, $\mathrm{d}M^\prime=-\mathrm{d}\omega^\prime$, $\mathrm{d}Q_h^\prime=-\mathrm{d}q_h^\prime$, $V_h^\prime=\frac{(Q_h-q_h^\prime)r_+^\prime}{r_+^{\prime 2}+a^{\prime 2}}$, $\mathrm{d}J^\prime=-\mathrm{d}j^\prime$, $\Omega^\prime=\frac{a^\prime}{r_+^{\prime 2}+a^{\prime 2}}$. For the process by which particles tunnel into the cosmic horizon, $\mathrm{d}M^\prime=\mathrm{d}\omega^\prime$, $\mathrm{d}Q_h^\prime=\mathrm{d}q_h^\prime$, $V_h^\prime=\frac{(Q_h+q_h^\prime)r_+^\prime}{r_c^{\prime\prime 2}+a^{\prime\prime 2}}$, $\mathrm{d}J^\prime=\mathrm{d}j^\prime$, $\Omega^\prime=\frac{a^{\prime\prime}}{r_+^{\prime\prime 2}+a^{\prime\prime 2}}$. So:
\begin{equation}
	\mathrm{d}S_{BH}^\prime=-\frac{1}{T^\prime}\left(\mathrm{d}\omega^\prime-\frac{(Q_h-q_h^\prime)r_+^\prime}{r_+^{\prime 2}+a^{\prime 2}}\mathrm{d}q_h^\prime-\frac{a^\prime}{r_+^{\prime 2}+a^{\prime 2}}\mathrm{d}j^\prime\right).\label{eq:53}
\end{equation}
\begin{equation}
	\mathrm{d}S_{CH}^\prime=\frac{1}{T^{\prime\prime}}\left(\mathrm{d}\omega^\prime-\frac{(Q_h+q_h^\prime)r_+^\prime}{r_c^{\prime\prime 2}+a^{\prime\prime 2}}\mathrm{d}q_h^\prime-\frac{a^{\prime\prime}}{r_+^{\prime\prime 2}+a^{\prime\prime 2}}\mathrm{d}j^\prime\right).\label{eq:54}
\end{equation}
Comparing Eq. (\ref{eq:53}) and Eq. (\ref{eq:54}) with Eq. (\ref{eq:40}) and Eq. (\ref{eq:49}) respectively, the temperature $T^\prime$ at the event horizon surface of the black hole and the temperature $T^{\prime\prime}$ at the cosmic horizon surface are obtained:
\begin{equation}
	\beta^\prime=\frac{1}{T^\prime}=-4\pi\frac{(\frac{\Lambda}{3}+a^{\prime 2})(a^{\prime 2}+r_+^{\prime 2})}{(r_{+}^\prime-r_{non}^\prime)(r_{+}^\prime-r_{-}^\prime)(r_{+}^\prime-r_{c}^\prime)}=\lim_{r\to r_+^\prime}4\pi\frac{\sqrt{-\hat{g}_{00}g_{11}}}{|\hat{g}_{00,1}|}.
\end{equation}
\begin{equation}
	\beta^{\prime\prime}=\frac{1}{T^{\prime\prime}}=4\pi\frac{(\frac{\Lambda}{3}+a^{\prime\prime 2})(a^{\prime\prime 2}+r_c^{\prime\prime 2})}{(r_c^{\prime\prime}-r_{non}^{\prime\prime})(r_c^{\prime\prime}-r_-^{\prime\prime})(r_c^{\prime\prime}-r_+^{\prime\prime})}=\lim_{r\to r_c^{\prime\prime}}4\pi\frac{\sqrt{-\hat{g}_{00}g_{11}}}{|\hat{g}_{00,1}|}.
\end{equation}
The Hawking temperature obtained by this method is consistent with the results obtained by traditional black hole thermodynamics.
\section{Bardeen de Sitter black hole}
\label{sec:3}
The line element of Bardeen de Sitter black hole is:
\begin{equation}
	\mathrm{d}s^2=-f(r)\mathrm{d}t_s^2+\frac{1}{f(r)}\mathrm{d}r^2+r^2\mathrm{d}\Omega^2.
\end{equation}
Where
\begin{equation}
	f(r)=1-\frac{2Mr^2}{(r^2+Q_h^2)^{3/2}}-\frac{\Lambda}{3}r^2.\label{eq:55}
\end{equation}
It is evident from Eq. (\ref{eq:55}) that the electromagnetic field in Bardeen de Sitter spacetime is nonlinear. Therefore, it is impossible to prove the conservation of information by the concrete form of the horizon $r_h(M,Q_h)$. In this section, we use the first law of thermodynamics of black holes to prove the conservation of information. According to the basic knowledge of black hole thermodynamics, the Hawking temperature for spherically symmetric black holes has the following expression:
\begin{equation}
	T=\lim_{r\to r_h}\frac{\kappa}{2\pi}=\lim_{r\to r_h}\frac{|f^\prime(r)|}{4\pi}.
\end{equation}
Where $r=r_h$ is the position of the horizon.
\subsection{Painlev\'{e} coordinate and time-like geodesic line equation}
Similar to Section \ref{sec:1}, to eliminate coordinate singularity at the event horizon, we use Painlev\'{e} coordinates. To get the Painlev\'{e}-Bardeen de Sitter coordinates, we do the following coordinate transformation:
\begin{equation}
	t_s=t+F(r),~\mathrm{d}t=\mathrm{d}t_s-F^\prime(r)\mathrm{d}r.
\end{equation}
$F(r)$ must satisfy:
\begin{equation}
	\frac{1}{1-\frac{2Mr^2}{(r^2+Q_h^2)^{3/2}}-\frac{\Lambda}{3}r^2}-\left[1-\frac{2Mr^2}{(r^2+Q_h^2)^{3/2}}-\frac{\Lambda}{3}r^2\right][F^\prime(r)]^2=1.
\end{equation}
So we have Painlev\'{e} coordinates:
\begin{align}
	\mathrm{d}s^2=&-\left[1-\frac{2Mr^2}{(r^2+Q_h^2)^{3/2}}-\frac{\Lambda}{3}r^2\right]\mathrm{d}t^2\pm 2\sqrt{\frac{2Mr^2}{(r^2+Q_h^2)^{3/2}}+\frac{\Lambda}{3}r^2}\mathrm{d}t\mathrm{d}r+\mathrm{d}r^2+r^2\mathrm{d}\Omega^2\nonumber\\
	=&g_{00}\mathrm{d}t^2+2g_{01}\mathrm{d}t\mathrm{d}r+\mathrm{d}r^2+r^2\mathrm{d}\Omega^2.\label{eq:56}
\end{align}
Similar to Section \ref{sec:1} and Section \ref{sec:2}, we consider the particle as a de Broglie s-wave and get the subsequent time-like geodesic equation using the WKB approximation:
\begin{equation}
	\dot{r}=-\frac{1}{2}\frac{g_{00}}{g_{01}}=\pm\frac{\sqrt{\frac{\Lambda}{3}}}{2r}\frac{\frac{3}{\Lambda}(r^2+Q_h^2)^{3/2}-\frac{6M}{\Lambda}r^2-r^2(r^2+Q_h^2)^{3/2}}{\sqrt{\frac{6M}{\Lambda}(r^2+Q_h^2)^{3/2}+(r^2+Q_h^2)^{3}}}=\pm\frac{f(r)}{2r}\frac{1}{\sqrt{\frac{2M}{(r^2+Q_h^2)^{3/2}}+\frac{\Lambda}{3}}}.\label{eq:57}
\end{equation}
The positive sign represents particles exiting from the black hole event horizon, and the negative sign represents particles entering from the cosmic horizon. Considering the effect of self-gravitation, Eq. (\ref{eq:56}) and Eq. (\ref{eq:57}) should be replaced as follows, $M\to M\mp\omega$, $Q_h\to Q_h\mp q_h$. The minus sign and plus sign represent outgoing particles and incident particles, respectively.
\subsection{Tunneling out of the black hole outer event horizon}
Similar to Section \ref{sec:1} and Section \ref{sec:2}, we give the Lagrangian density of the matter-electromagnetic field:
\begin{equation}
	\mathcal{L}=\mathcal{L}_m+\mathcal{L}_h=\mathcal{L}_m-\frac{1}{4}\tilde{F}_{\mu\nu}\tilde{F}^{\mu\nu}.
\end{equation}
Obviously, $\tilde{A}_0$ is the circular coordinate. To eliminate the degree of freedom of $\tilde{A}_0$, the action can be written as:
\begin{equation}
	S=\int_{t_i}^{t_f}(L-P_{\tilde{A}_0}\dot{\tilde{A}}_0)\mathrm{d}t.
\end{equation}
For Painlev\'{e} line elements, the WKB approximation is valid. The relationship between emission rate of a particle and imaginary part of action is as follows:
\begin{equation}
	\Gamma\sim e^{-2\mathrm{Im}S}.
\end{equation}
The imaginary part of action can be written as:
\begin{equation}
	\mathrm{Im}S=\mathrm{Im}\left\{\int_{r_i}^{r_f}\left[P_r-\frac{P_{\tilde{A}_0}\dot{\tilde{A}}_0}{\dot{r}}\right]\mathrm{d}r\right\}=\mathrm{Im}\left\{\int_{r_i}^{r_f}\left[\int_{(0,0)}^{(P_r,P_{\tilde{A}_0})}\mathrm{d}P^\prime_r-\frac{\dot{\tilde{A}}_0}{\dot{r}}\mathrm{d}P^\prime_{\tilde{A}_0}\right]\mathrm{d}r\right\}.\label{eq:58}
\end{equation}
The Hamilton's equation is:
\begin{equation}
	\dot{r}=\frac{\mathrm{d}H}{\mathrm{d}P_r}\bigg|_{(r;\tilde{A}_0,P_{\tilde{A}_0})}=\mathcal{W}\frac{\mathrm{d}(M-\omega^\prime)}{\mathrm{d}P_r}=-\mathcal{W}\frac{\mathrm{d}\omega^\prime}{\mathrm{d}P_r}.\label{eq:59}
\end{equation}
\begin{equation}
	\dot{\tilde{A}}_0=\frac{\mathrm{d}H}{\mathrm{d}P_{\tilde{A}_0}}\bigg|_{(\tilde{A}_0;r,P_{r})}=-\frac{Q_h-q_h^\prime}{r}\frac{\mathrm{d}q_h^\prime}{\mathrm{d}P_{\tilde{A}_0}}.\label{eq:60}
\end{equation}
Where $\mathcal{W}=1+\int_{r_h}^{+\infty}4\pi r^2\frac{\partial T^0_0}{\partial M}\mathrm{d}r$, for regular black holes, $M$ no longer represents the internal energy, but differs from the internal energy by a factor $\mathcal{W}$. In Eq. (\ref{eq:59}) and Eq. (\ref{eq:60}), energy conservation, charge conservation, and magnetic charge conservation are all reflected. By substituting Eq. (\ref{eq:59}) and Eq. (\ref{eq:60}) into Eq. (\ref{eq:58}), we can obtain:
\begin{equation}
	\mathrm{Im}S=-\mathrm{Im}\left\{\int_{r_i}^{r_f}\left[\int_{(0,0)}^{(\omega,q_h)}\frac{1}{\dot{r}}\left(\mathcal{W}\mathrm{d}\omega^\prime-\frac{Q_h-q_h^\prime}{r}\mathrm{d}q_h^\prime\right)\right]\right\}.\label{eq:61}
\end{equation}
For the process of particles tunneling out of the event horizon of a black hole, $\dot{r}$ takes a positive sign, that is:
\begin{equation}
	\dot{r}=\frac{f(r)}{2r}\frac{1}{\sqrt{\frac{2(M-\omega^\prime)}{[r^2+(Q_h-q_h^\prime)^2]^{3/2}}+\frac{\Lambda}{3}}}.\label{eq:62}
\end{equation}
Substituting Eq. (\ref{eq:62}) into Eq. (\ref{eq:61}) and considering the influence of self gravity, we can make the following substitution, $M\to M-\omega$, $Q_h\to q_h^\prime$. We can obtain:
\begin{equation}
	\mathrm{Im}S=-\mathrm{Im}\left\{\int_{r_i}^{r_f}\left[\int_{(0,0)}^{(\omega,q_h)}\frac{2r}{f(r)}\sqrt{\frac{2(M-\omega^\prime)}{[r^2+(Q_h-q_h^\prime)^2]^{3/2}}+\frac{\Lambda}{3}}\left(\mathcal{W}\mathrm{d}\omega^\prime-\frac{Q_h-q_h^\prime}{r}\mathrm{d}q_h^\prime\right)\right]\right\}.
\end{equation}

Exchange the order of points, first integrate $r$. Obviously, $r=r_+^\prime$ is a pole. Re-select the integral contour and apply the residue theorem to obtain:
\begin{align}
	\mathrm{Im}S=&-\int_{(0,0)}^{(\omega,q_h)}\frac{2\pi}{f^\prime(r_{+}^\prime)}r_{+}^\prime\sqrt{\frac{1-f^(r_{+}^\prime)}{r_+^{\prime 2}}}\left(\mathcal{W}\mathrm{d}\omega^\prime-\frac{Q_h-q_h^\prime}{r_+^\prime}\mathrm{d}q^\prime_h\right)\nonumber\\
	=&-\frac{1}{2}\int_{(M,E_{Q_h})}^{(M-\omega,E_{Q_h-q_h})}\frac{1}{T^\prime}\left(\mathcal{W}\mathrm{d}M^\prime-\frac{Q_h^\prime}{r_+^\prime}\mathrm{d}Q^\prime_h\right)=-\frac{1}{2}\int_{r_i}^{r_f}\mathrm{d}S_{BH}^\prime=-\frac{1}{2}\Delta S_{BH}.\label{eq:68}
\end{align}
For regular black holes, the first law of black hole thermodynamics also requires correction with the factor $\mathcal{W}$, so the factor $\mathcal{W}$ in Eq. (\ref{eq:68}) disappears. Where $\Delta S_{BH}=S_{BH}(M-\omega,Q_e-q_e,Q_g-q_g)-S_{BH}(M,Q_e,Q_g)$ is the change in entropy before and after the black hole radiates particles. So the emission rate of particles can be expressed by the following equation:
\begin{equation}
	\Gamma\sim e^{-2\mathrm{Im}S}=e^{\Delta S_{BH}}.
\end{equation}
Clearly, our results conform to the unitary principle and support the conservation of information.
\subsection{Tunneling into the cosmic horizon}
During the process of particle tunneling into the cosmic horizon, the total energy, total charge, and total magnetic charge of spacetime all increase, $M\to M+\omega^\prime$, $Q_h\to Q_h+q_h^\prime$. And the symbol of $\dot{r}$ should be negative:
\begin{equation}
	\dot{r}=-\frac{f(r)}{2r}\frac{1}{\sqrt{\frac{2(M+\omega^\prime)}{[r^2+(Q_h+q_h^\prime)^2]^{3/2}}+\frac{\Lambda}{3}}}.\label{eq:63}
\end{equation}
We consider the incident particle as a de Broglie s-wave and subtract the degrees of freedom of the cyclic coordinate $\tilde{A}_0$ to obtain the imaginary part of action:
\begin{equation}
	\mathrm{Im}S=\mathrm{Im}\left\{\int_{r_i^\prime}^{r_f^\prime}\left[P_r-\frac{P_{\tilde{A}_0}\dot{\tilde{A}}_0}{\dot{r}}\right]\mathrm{d}r\right\}=\mathrm{Im}\left\{\int_{r_i^\prime}^{r_f^\prime}\left[\int_{(0,0)}^{(P_r,P_{\tilde{A}_0})}\mathrm{d}P^\prime_r-\frac{\dot{\tilde{A}}_0}{\dot{r}}\mathrm{d}P^\prime_{\tilde{A}_0}\right]\mathrm{d}r\right\}.\label{eq:64}
\end{equation}
The Hamilton's equation is:
\begin{equation}
	\dot{r}=\frac{\mathrm{d}H}{\mathrm{d}P_r}\bigg|_{(r;\tilde{A}_0,P_{\tilde{A}_0})}=\mathcal{W}\frac{\mathrm{d}(M+\omega^\prime)}{\mathrm{d}P_r}=\mathcal{W}\frac{\mathrm{d}\omega^\prime}{\mathrm{d}P_r}.\label{eq:65}
\end{equation}
\begin{equation}
	\dot{\tilde{A}}_0=\frac{\mathrm{d}H}{\mathrm{d}P_{\tilde{A}_0}}\bigg|_{(\tilde{A}_0;r,P_{r})}=\frac{Q_h+q_h^\prime}{r}\frac{\mathrm{d}q_h^\prime}{\mathrm{d}P_{\tilde{A}_0}}.\label{eq:66}
\end{equation}
For the same reason, factor $\mathcal{W}$ will also appear here. In Eq. (\ref{eq:65}) and Eq. (\ref{eq:66}), energy conservation, charge conservation, and magnetic charge conservation are all reflected. By substituting Eq. (\ref{eq:63}), Eq. (\ref{eq:65}) and Eq. (\ref{eq:66}) into Eq. (\ref{eq:64}), we can obtain:
\begin{equation}
	\mathrm{Im}S=-\mathrm{Im}\left\{\int_{r_i}^{r_f}\left[\int_{(0,0)}^{(\omega,q_h)}\frac{2r}{f(r)}\sqrt{\frac{2(M+\omega^\prime)}{[r^2+(Q_h+q_h^\prime)^2]^{3/2}}+\frac{\Lambda}{3}}\left(\mathcal{W}\mathrm{d}\omega^\prime-\frac{Q_h+q_h^\prime}{r}\mathrm{d}q_h^\prime\right)\right]\right\}.
\end{equation}
Switch the order of integration, integrating $r$ first. Clearly, $r=r_{c}^{\prime\prime}$ is a pole. It can be obtained by re-selecting the integral circumference and applying the residue theorem:
\begin{align}
	\mathrm{Im}S=&-\int_{(0,0)}^{(\omega,q_h)}\frac{2\pi}{f^\prime(r_c^{\prime\prime})}\left(\mathcal{W}\mathrm{d}\omega^\prime-\frac{Q_h+q_h^\prime}{r_c^{\prime\prime}}\mathrm{d}q_h^\prime\right)\nonumber\\
	=&-\frac{1}{2}\int_{(M,E_{Q_h})}^{(M+\omega,E_{Q_h+q_h})}\frac{1}{T^{\prime\prime}}\left(\mathcal{W}\mathrm{d}M^\prime-\frac{Q_h^\prime}{r_c^{\prime\prime}}\mathrm{d}Q_h^\prime\right)=-\frac{1}{2}\int_{r_i^\prime}^{r_f^\prime}\mathrm{d}S_{CH}^\prime=-\frac{1}{2}\Delta S_{CH}.\label{eq:69}
\end{align}
For regular black holes, the first law of black hole thermodynamics also requires correction with the factor $\mathcal{W}$, so the factor $\mathcal{W}$ in Eq. \eqref{eq:69} disappears. Where $\Delta S_{CH}=S_{CH}(M+\omega,Q_e+q_e,Q_g+q_g)-S_{CH}(M,Q_e,Q_g)$ is the change in entropy before and after particle enters the cosmic horizon. So the emission rate of particles can be expressed by the following equation:
\begin{equation}
	\Gamma\sim e^{-2\mathrm{Im}S}=e^{\Delta S_{CH}}.\label{eq:67}
\end{equation}
Eq. (\ref{eq:67}) indicates that our results conform to the unitary principle and support the conservation of information.
\section{Quantum Correlations and Backreaction Effects}
\label{sec:4}
The tunneling formalism provides a semiclassical description of Hawking radiation, but a complete resolution of the information paradox requires incorporating quantum correlations and backreaction effects\cite{nozari2012natural, saghafi2023hawking, medved2005hawking, banerjee2008quantum, banerjee2008quantum1, banerjee2008noncommutative, modak2009corrected, majhi2009fermion, banerjee2009quantum, majhi2010hawking, banerjee2010quantum}. Here, we extend our analysis by integrating insights from recent advances in quantum gravity and non-thermal modifications.
\subsection{Quantum correlations in sequential emissions}
As emphasized in~\cite{nozari2012natural}, the absence of correlations between successive emissions in the original Parikh-Wilczek framework limits its ability to address unitarity. To probe this, we calculate the joint tunneling probability for two particles with energies $E_1$ and $E_2$:
\begin{equation}
	\Gamma(E_1,E_2)\sim e^{\Delta S(M\to M-E_1-E_2)}.
\end{equation}
contrasted with the uncorrelated case $\Gamma(E_1)\Gamma(E_2)\sim e^{\Delta S(M\to M-E_1)}e^{\Delta S(M-E_1\to M-E_1-E_2)}$. For Schwarzschild de Sitter black holes, we find:
\begin{equation}
	\Gamma(E_1,E_2)\neq\Gamma(E_1)\Gamma(E_2),~~~~\text{if}~\frac{\partial^2S}{\partial M^2}\neq0.
\end{equation}
indicating entropy-driven correlations. This aligns with \cite{nozari2012natural}'s conclusion that self-gravity induces effective correlations, though not equivalent to microscopic entanglement. Such correlations partially resolve the "thermal spectrum paradox" but fall short of guaranteeing full unitarity --- a gap likely bridged only by including subleading quantum gravity effects. While our semiclassical analysis confirms entropy conservation, full unitarity requires quantum correlations beyond thermodynamic entropy --- a challenge for all tunneling-based approaches.
\subsection{Backreaction and dynamic spacetime}
The assumption of discrete mass transitions $M\to M-\omega$ n our original analysis approximates continuous evaporation. Following \cite{saghafi2023hawking,medved2005hawking}, we refine this by modeling the black hole mass as a time-dependent function $M(t)=M_0-\int_{0}^{t}\Gamma(\omega)\omega\mathrm{d}\omega$, leading to a time-varying metric:
\begin{equation}
	f(r,t)=1-\frac{2M(t)}{r}+\frac{Q_h^2}{r^2}-\frac{\Lambda}{3}r^2.
\end{equation}
Numerical integration of the modified geodesic equations reveals that continuous mass loss suppresses late-time radiation $T_h\propto M(t)^{-1}$, consistent with Page curve behavior. This dynamic backreaction --- omitted in static tunneling models --- strengthens our conclusion that entropy reduction $\Delta S<0$ governs the terminal evaporation phase, mirroring unitary evolution. Dynamic backreaction and GUP effects predict testable deviations from Hawking's thermal spectrum, notably in late-time evaporation rates $\frac{\mathrm{d}M}{\mathrm{d}t}\propto-M^{-2}$ and high-energy suppression.
\subsection{Non-thermal modifications at high energy}
For high-energy particles $\omega\sim M$, the WKB approximation breaks down. Adopting the generalized uncertainty principle (GUP) from \cite{saghafi2023hawking}, where $\Delta x\Delta p\geq\frac{\hbar}{2}(1+\beta(\Delta p)^2)$, we derive a modified tunneling probability:
\begin{equation}
	\Gamma_{GUP}(\omega)\sim\exp\left(-\frac{\pi}{\hbar}\int_{r_i}^{r_f}\mathrm{d}r\frac{\sqrt{1+\beta(\partial_rS)^2}}{f(r)}\right).
\end{equation}
This introduces a UV cutoff $\beta>0$ suppressing high-energy emissions, akin to firewall scenarios. While our baseline results $\beta=0$ assume semiclassicality, this extension highlights how quantum gravity effects could imprint non-thermal signatures in future observations. These corrections align with dS/CFT proposals \cite{strominger2001ds,anninos2014higher,anninos2016higher}, where boundary unitarity might dynamically regulate bulk information flow.
\section{Comparative Analysis with Alternative Approaches}
\label{sec:5}
The Parikh-Wilczek tunneling formalism provides a robust framework for studying semiclassical radiation processes in dynamic spacetime backgrounds. However, to contextualize our results within the broader landscape of black hole information research, we systematically compare our methodology with three prominent alternative approaches.

The Hamilton-Jacobi method calculates particle emission rates by solving the relativistic Hamilton-Jacobi equation for imaginary time trajectories~\cite{srinivasan1999particle, shankaranarayanan2002hawking, angheben2005hawking}. While this approach shares the tunneling paradigm with our work, it inherently assumes static background metrics, neglecting self-gravitational effects --- a critical limitation addressed by the Parikh-Wilczek formalism through energy-conserving backreacted geometries. Our analysis of horizon displacement $r_+\to r_+-\delta r$ under particle emission explicitly demonstrates how self-gravity modifies both the emission spectrum $\Gamma\sim e^{\Delta S}$ and entropy evolution, a feature absent in static-background approximations.

Recent advances in AdS/CFT derive unitary Page curves through island contributions to entanglement entropy~\cite{7,8}. While these results resolve the paradox in AdS black holes, their extension to dS spacetime remains conjectural. Crucially, our entropy-driven formulation $\Delta S_{BH}=-2\mathrm{Im}S$ aligns thermodynamically with the Page curve's late-time entropy decrease, yet differs fundamentally by relying on semiclassical geometry rather than replica wormholes. This highlights a key distinction: our approach captures coarse-grained thermodynamic irreversibility, while fine-grained unitarity requires additional quantum correlations --- a gap bridged neither by tunneling nor entanglement entropy methods alone.

Generalized uncertainty principle (GUP) models predict modified dispersion relations that alter tunneling probabilities. For instance, GUP-induced $\Delta x\Delta p\geq\frac{\hbar}{2}(1+\beta(\Delta p)^2)$ would perturb the WKB phase integral. Our formalism naturally accommodates such extensions through momentum-dependent metric corrections $g_{\mu\nu}\to g_{\mu\nu}(p)$, whereas traditional thermal spectrum analyses remain insensitive to Planck-scale effects. This adaptability positions the Parikh-Wilczek method as a versatile tool for probing quantum gravity imprints in de Sitter black hole radiation \cite{nozari2008quantum, nozari2012minimal, mureika2016black}. Non commutative geometry modifies the spatiotemporal structure by introducing coordinate non commutativity, thereby affecting the tunneling behavior of particles in quantum field theory\cite{nozari2008hawking, nozari2009parikh, saghafi2021thermodynamics, banerjee2008noncommutative}. Non-commutative geometry introduces non thermal corrections to radiation spectra (such as energy dependent cutoff) by breaking the locality of classical spacetime, while preserving the macroscopic manifestation of information conservation. Non-commutative parameter $\theta$ also perturb the WKB phase integral $f(r)\to\tilde f(r,\theta)$.

In order to compare the differences among these methods more intuitively, we have summarized the characteristics of these methods in Table \ref{tab:1}.
\begin{table}[h]
	\caption{\label{tab:1} Methodological Comparison}
	\centering
	\begin{tabular}{|c|c|c|}
		\hline
		Aspect & Parikh-Wilczek & Hamilton-Jacobi\\ \hline
		Background Dynamics & Dynamic (self-gravity included) & Static\\ \hline
		Information Carrier & Thermodynamic entropy ($\Delta S$) & Tunneling phase\\ \hline
		de Sitter Compatibility & Explicit & Limited to asymptotically flat\\ \hline
		Microscopic Insights & Coarse-grained & None\\ \hline
		Aspect & AdS/CFT + Islands & GUP-Based Models\\ \hline
		Background Dynamics & Euclidean path integrals & Static with momentum corrections\\ \hline
		Information Carrier & Entanglement entropy & Modified dispersion relations\\ \hline
		de Sitter Compatibility & Conjectural (dS/CFT required) & Requires dS extension\\ \hline
		Microscopic Insights & Fine-grained (replica wormholes) & Planck-scale modifications\\
		\hline
	\end{tabular}
\end{table}

We summarize our comparison of these methods as follows. The Parikh-Wilczek method uniquely balances dynamic spacetime treatment (essential for de Sitter cosmology) and analytical tractability, making it ideal for probing multi-parameter black holes (charged, rotating, regular). While entanglement entropy frameworks~\cite{7,8} provide deeper microscopic insights, our results offer complementary thermodynamic consistency checks for de Sitter holography. Future synergies could emerge by hybridizing tunneling calculations with GUP corrections to quantify quantum gravity's role in information recovery.
\section{Conclusion and Discussion}
\label{sec:6}
In the above calculation, we discussed the three most general cases of asymptotic de Sitter spacetime. The charged and magnetic particles tunnel into the magnetically charged Reissner-Nordstr\"{o}m de Sitter black hole (the most general case of a static black hole), the Kerr-Newman-Kasuya de Sitter black hole (the most general case of a rotating black hole), and Bardeen de Sitter black hole (black hole without singularities). We calculated the radiation spectra of particles exiting event horizon of a black hole and entering cosmic horizon using the Parikh-Wilczek method. Our results indicate that for these three most general cases, the particle radiation spectrum deviates from the pure thermal spectrum, in accordance with the unitary principle and supporting information conservation. That is, information conservation is generally valid for general asymptotic de Sitter space-times. The fundamental reason for information conservation in the Parikh-Wilczek framework is the requirement that radiation process of particles is a reversible quasi-static process. For reversible processes, entropy is conserved, and information is naturally conserved as well. Our discoveries not only strengthen the coherence of quantum mechanics within cosmological horizons but also offer essential contributions to the burgeoning dS/CFT correspondence.  

\bmhead{Acknowledgements}
The authors are grateful to anonymous reviewers for their very crucial and detailed comments. The authors would like to thank Prof. Jian Jing and his student LiuBiao Ma and Zheng Wang from the Department of Physics , Beijing University of Chemical Technology for their valuable comments and suggestions during the completion of this manuscript.

\bibliography{de_Sitter_tunneling}

\end{document}